\documentclass[twocolumn]{article}
\usepackage{cite}
\usepackage{amsmath,amssymb,amsfonts}

\usepackage{graphicx}
\usepackage{hyperref}
\usepackage{textcomp}
\def\BibTeX{{\rm B\kern-.05em{\sc i\kern-.025em b}\kern-.08em
    T\kern-.1667em\lower.7ex\hbox{E}\kern-.125emX}}
\usepackage{amsthm}

\providecommand{\keywords}[1]
{
	\small	
	\textbf{\textit{Keywords---}} #1
}

\usepackage{graphicx, color, caption, subcaption}
\graphicspath{{fig/}}

\usepackage{pgf}
\usepackage{bm}

\newtheorem{thm}{Theorem}
\newtheorem{lem}{Lemma}
\newtheorem{prob}{Problem}
\newtheorem{cor}[thm]{Corollary}

\newtheorem{definition}{Definition}
\newtheorem{assumption}{Assumption}

\newcommand{\guillemets}[1]{``#1''}
\newcommand{\set}[1]{\left\{#1\right\}}
\newcommand{\case}[1]{\emph{\underline{#1}:}}

\newcommand{\myvector}[1]{\bm{#1}}
\newcommand{\myrand}[1]{\underline{#1}}
\newcommand{\myrandvector}[1]{\myrand{\myvector{#1}}}
\newcommand{\mymatrix}[1]{\bm{#1}}

\newcommand{\cent}{{\mathrm{c}}}

\newcommand{\N}{\mathbb{N}} 
\newcommand{\R}{\mathbb{R}} 
\newcommand{\E}{\mathrm{E}} 

\newcommand{\eqdef}{\triangleq}

\newcommand{\rA}{\mathcal{A}}

\newcommand{\rC}{\mathcal{C}}

\newcommand{\rE}{\mathcal{E}}

\newcommand{\rK}{\mathcal{K}}

\newcommand{\rN}{\mathcal{N}}
\newcommand{\rP}{\mathcal{P}}

\newcommand{\rV}{\mathcal{V}}

\newcommand{\intEnt}[2]{\set{#1, \dots, #2}}
\newcommand{\norm}[1]{\left\lVert#1\right\rVert}

\newcommand{\suchthat}{\ | \ }
\newcommand{\ie}{{i.e.},}
\newcommand{\eg}{{e.g.},}

\newcommand{\diag}{\mathrm{diag}}
\DeclareMathOperator*{\trace}{tr}
\DeclareMathOperator*{\interior}{int}
\DeclareMathOperator*{\minimize}{minimize}
\DeclareMathOperator*{\subject}{subject\, to:}
\DeclareUnicodeCharacter{2212}{-}

\newcommand{\com}[1]{#1}

\begin{document}
\title{Revisiting Split Covariance Intersection: Correlated Components and Optimality}
\author{Colin Cros, Pierre-Olivier Amblard, Christophe Prieur, Jean-François Da Rocha
\thanks{C. Cros is with the ISAE-SUPAERO, Univ. Toulouse, France. P.-O. Amblard and C. Prieur are with the CNRS, Univ. Grenoble Alpes, GIPSA-lab, F-38000 Grenoble, Auvergne-Rhône-Alpes, France. J.-F. Da Rocha is with Telespazio FRANCE, F-31100 Toulouse, Occitanie, France.
When this work was conducted, C. Cros was with the CNRS and Telespazio FRANCE.
Corresponding author: {\tt\small colin.cros@isae.fr.}
}
}

\maketitle
\begin{abstract}
	Linear fusion is a cornerstone of estimation theory. Implementing optimal linear fusion requires knowledge of the covariance of the vector of errors associated with all the estimators. In distributed or cooperative systems, the cross-covariance terms cannot be computed, and to avoid underestimating the estimation error, conservative fusions must be performed. A conservative fusion provides a fused estimator with a covariance bound that is guaranteed to be larger than the true, but computationally intractable, covariance of the error. Previous research by Reinhardt \textit{et al.} proved that, if no additional assumption is made about the errors of the estimators, the minimal bound for fusing two estimators is given by a fusion called Covariance Intersection (CI). In distributed systems, the estimation errors contain independent and correlated terms induced by the measurement noises and the process noise. In this case, CI is no longer the optimal method. Split Covariance Intersection (SCI) has been developed to take advantage of the uncorrelated components. This paper extends SCI to also take advantage of the correlated components. Then, it is proved that the new fusion provides the optimal conservative fusion bounds for two estimators, generalizing the optimality of CI to a wider class of fusion schemes. The benefits of this extension are demonstrated in simulations.
\end{abstract}


\section{Introduction}

In a sensor network, the estimation of the target state can be carried out centrally or on a distributed basis. In centralized networks, the sensors send their measurements to a computing station, which performs the estimation. In contrast, in distributed networks, the sensors process the measurements and perform the estimation themselves, sharing certain information with their neighbors. There exist many variations between these two extremes, see \eg{} \cite{he2020distributed} for a detailed study. Centralized approaches generally provide better estimates by optimally fusing the sensor measurements. However, they are prone to failure, and require high communication and computation costs. On the other hand, distributed networks are more robust and are easily scalable. However, since the nodes have only local information, the estimation algorithms should be carefully designed to avoid information redundancy that leads to overoptimistic estimates. This paper focuses on distributed networks.

In a typical distributed algorithm, each node has an estimator of the target state. At each time step, the nodes carry out measurements, update their estimates, and share them with their neighbors in the network. Each node must then combine the estimates received from its neighbors with its own; this process is known as fusion.
The problem of optimal fusion has been studied for decades \cite{bar1986effect}, and especially when the fused estimator is sought as a linear combination of the estimators. The optimal linear fusion of two estimators was first proposed by Bar-Shalom and Campo \cite{bar1986effect}. They emphasized the importance of the cross-covariance term. Since then, extensions have been derived for the fusion of any number of estimators, see \eg{} \cite{li2003optimal}. To perform the optimal linear fusion, full knowledge of the second-order moments of the estimation errors is required, \ie{} the covariance matrix associated with the centralized vector of errors must be known. In particular, knowledge of the covariances of each estimator is not sufficient, but knowledge of the cross-covariances between each pair of estimators is also required. This requirement may be prohibitive for applications such as distributed estimation, where the nodes have access only to local information.
Several strategies have been proposed to perform the fusion when the cross-covariances are unknown and cannot be computed. The simplest is to assume that the estimators are uncorrelated, and then apply the optimal scheme. However, this naive strategy leads to an underestimation of the estimation error, see \eg{} \cite{arambel2001covariance}, and should therefore be avoided. Furthermore, as \cite{uhlmann2003covariance} points out, selecting any particular possible centralized covariance to apply the optimal scheme also leads to an underestimation of the error. As a consequence, the whole set of \emph{admissible} centralized covariances should be considered when designing the fusion. The covariance of the error of the resulting fused estimator cannot be computed since each centralized covariance would produce a different fused covariance. Instead, a \emph{conservative} bound is sought to ensure that the estimation error is not underestimated. This bound should be larger than the covariance of the estimation error of the fused estimator for all admissible covariances. The optimal linear fusion problem then consists in finding a fused estimator with the smallest conservative bound. This problem can be formulated as a general optimization problem \cite{forsling2022conservative}.
\com{Covariance Intersection (CI) \cite{julier1997nondivergent} fuses estimators without information about their cross-covariances. It gives the optimal bound \cite{reinhardt2015minimum} when fusing two estimators. However, the set of admissible covariances considered by CI is often too conservative for practical applications. Better fusion schemes have been proposed when additional assumptions can be made. For example, Partitioned CI (PCI) \cite{ajgl2019rectification, ajgl2022covariance} proposes tighter bounds than CI when one or several blocks of the covariance of the centralized error vector are unknown. Inverse CI (ICI) provides better conservative bounds than CI \cite{noack2017decentralized} when the estimators are assumed to share a common estimate. Under some additional assumptions, ICI also applies when the estimators share unknown correlated components \cite{noack2017inverse}. Finally, when the estimation errors are known to have uncorrelated components (generally independent), Split CI (SCI) provides a better result \cite{julier2001general}.}
Such independent terms in the errors are commonly encountered in practice, for example, when measurement errors between senors are independent of each other. SCI has been applied to a variety of problems, such as SLAM \cite{julier2001simultaneous, julier2007using}, cooperative localization \cite{li2013cooperative, carrillo2013decentralized}, or cooperative perception \cite{lima2021data}. However, SCI has a shortcoming: it can only handle uncorrelated components. When the estimates are all corrupted by a common error, for example a process noise, SCI cannot exploit this correlated term. Furthermore, despite SCI's excellent practical performance, to the best of our knowledge, no theoretical performance studies have ever been carried out.

The goal of this paper is twofold. First, SCI is improved to exploit not only the uncorrelated error components, but also the correlated ones. This new fusion, called Extended SCI (ESCI), unifies several commonly used fusions under a single formalism and extends our preliminary work presented in \cite{cros2024split}. Second, the performances of this fusion are justified by a theoretical result: ESCI provides the optimal conservative fusion bound for the fusion of two estimators. This result is a generalization of the work of Reinhardt \emph{et al.} \cite{reinhardt2015minimum} who proved the equivalent result in the context of CI. In \cite{reinhardt2015minimum}, the authors used the fact that conservative bounds must contain a minimal volume corresponding, in the context of CI, to an intersection of ellipsoids. They used a result of Kahan \cite{kahan1968circumscribing} on the characterization of the intersection of two ellipsoids. In the context of ESCI, this minimal volume has no special structure and Kahan's result cannot be used. An equivalent result must be formally redeveloped in the context of ESCI.
This result justifies the use of several commonly used fusions. Furthermore, the techniques used in the proof provide an understanding of why this fusion is optimal and may help in the design of new fusions for more than two estimators.

\begin{figure*}
	\centering
	\null\hfill
	\subfloat[CI bounds.\label{sfig: CI bounds}]{\input{fig/fusion_ic.pgf}}
	\hfill
	\subfloat[SCI bounds.\label{sfig: SCI bounds}]{\input{fig/fusion_icf.pgf}}
	\hfill
	\subfloat[ESCI bounds.\label{sfig: ESCI bounds}]{\input{fig/fusion_icfg.pgf}}
	\hfill\null
	\caption{Comparison of the CI, SCI and ESCI bounds for the fusion of two estimators whose errors are decomposed as \eqref{eq: Example spliting update SCI}. The dashed lines represent the covariances $\mymatrix{\tilde P}_1$ (blue) and $\mymatrix{\tilde P}_2$ (red). The grey solid lines represent the conservative upper bounds obtained with the different methods and the vectors $\myvector{\omega_i} = \begin{pmatrix} i/5 & 1 - i/5\end{pmatrix}^{\intercal}$ for $i\in\intEnt{0}{5}$. The numerical values are provided at the end of Section~\ref{ssec: Motivating example}.} 
	\label{fig: Comparison of the bounds}
\end{figure*}

The rest of the paper is organized as follows. Section~\ref{sec: Background} introduces linear fusions, the notion of conservatism, and SCI. Then, the new ESCI is introduced in Section~\ref{sec: ESCI}. The main result is given and proved in Section~\ref{sec: Optimality}. Section~\ref{sec: Application} demonstrates the advantages of the new ESCI over SCI in a distributed estimation context. A discussion is proposed in Section~\ref{sec: Discussion}. Finally, Section~\ref{sec: Conclusion} gives some perspectives. To lighten the reading load, the proofs of the lemmas are given in the appendices.

\bigbreak
\textbf{Notation.}
In the sequel, $d$ denotes the dimension of the state and $N$ the number of sensors. Vectors are denoted in lowercase boldface letters \eg{} $\myvector{x} \in \R^d$, and matrices in uppercase boldface variables \eg{} $\mymatrix{M} \in \R^{d\times d}$. The vectors of $\R^{Nd}$ are denoted in typewritter style \eg{} $\myvector{\mathtt{x}}_\cent \in \R^{Nd}$. Random variables are underlined \eg{} $\myrandvector{x}$ for a random vector. The notation $\E[\cdot]$ denotes the expected value of a random variable and $\norm{\cdot}$ the Euclidean norm of a vector. The trace, the inverse and the transpose of a matrix $\mymatrix{M}$ and the identity matrix are denoted as $\trace \mymatrix{M}$, $\mymatrix{M}^{-1}$, $\mymatrix{M}^\intercal$ and $\mymatrix{I}$ respectively. For two matrices $\mymatrix{A}$ and $\mymatrix{B}$, the notations $\mymatrix{A} \preceq \mymatrix{B}$ and $\mymatrix{A} \prec \mymatrix{B}$ mean that the difference $\mymatrix{B} - \mymatrix{A}$ is positive semi-definite and positive definite respectively. The notation $\mymatrix{A} \otimes \mymatrix{B}$ denotes the Kronecker product of $\mymatrix{A}$ and $\mymatrix{B}$. For a positive semi-definite matrix $\mymatrix{A}$, $\mymatrix{A}^{1/2}$ \com{denotes its symmetric positive semi-definite} square roots. The unit simplex of $\R^N$ is denoted as $\rK^N \eqdef \set{\myvector{\omega}\in \R^N \suchthat \mymatrix{1}_N^{\intercal} \myvector{\omega} =1,\ \omega_i \ge 0}$ and its interior $\interior \rK^N$.
A positive definite matrix $\mymatrix{P}$ is represented in the figures by the ellipsoid $\rE(\mymatrix{P}) \eqdef \set{\myvector{x} \suchthat \myvector{x}^\intercal \mymatrix{P}^{-1}\myvector{x} \le 1}$. The inequality $\mymatrix{P} \preceq \mymatrix{Q}$ is geometrically equivalent to $\rE(\mymatrix{P}) \subseteq \rE(\mymatrix{Q})$.

\section{Background}\label{sec: Background}

\subsection{Linear fusion}
Consider a random state $\myrandvector{x} \in \R^d$ and $N \ge 2$ unbiased estimators of $\myrandvector{x}$ denoted as $\myrandvector{\hat x}_i$ for $i \in \intEnt{1}{N}$. The estimation errors are denoted as $\myrandvector{\tilde x}_i \eqdef \myrandvector{\hat x}_i - \myrandvector{x}$, and their covariances and cross-covariances as $\mymatrix{\tilde P}_i \eqdef \E[\myrandvector{\tilde x}_i \myrandvector{\tilde x}_i^\intercal]$ and $\mymatrix{\tilde P}_{i,j} \eqdef \E[\myrandvector{\tilde x}_i \myrandvector{\tilde x}_j^\intercal]$. A linear fusion consists in creating a new unbiased estimator $\myrandvector{\hat x}_F$ ($F$ for \emph{fused}) as a linear combination of the $\myrandvector{\hat x}_i$. It is defined by a gain matrix $\mymatrix{K} = \begin{bmatrix}
\mymatrix{K}_1 & \cdots & \mymatrix{K}_N \end{bmatrix} \in \R^{d \times Nd}$ as:
\begin{equation}\label{eq: Fused state}
	\myrandvector{\hat x}_F(\mymatrix{K}) \eqdef \sum_{i = 1}^N \mymatrix{K}_i \myrandvector{\hat x}_i = \mymatrix{K} \myrandvector{\hat \mathtt{x}}_\cent,
\end{equation}
where $\myrandvector{\hat \mathtt{x}}_\cent \eqdef \begin{pmatrix} \myrandvector{\hat x}_1^{\intercal} & \cdots & \myrandvector{\hat x}_N^{\intercal} \end{pmatrix}^{\intercal} \in \R^{Nd}$ is the centralized vector of estimators. Introduce the centralized observation matrix $\mymatrix{H} \eqdef \myvector{1}_N \otimes \mymatrix{I}_d \in \R^{Nd \times d}$, the unbiasedness of $\myrandvector{\hat x}_F(\mymatrix{K})$ imposes that the gain must satisfy:
\begin{equation}\label{eq: Constraint unbiasedness}
	\mymatrix{K} \mymatrix{H} = \mymatrix{I}_d.
\end{equation}
When there is no ambiguity on the gain, we suppress the dependence in $\mymatrix{K}$ and denote the estimator as $\myrandvector{\hat x}_F$.
The error of the fused estimator and its covariance are:
\begin{subequations}
	\begin{align}
		\myrandvector{\tilde x}_F &\eqdef \myrandvector{\hat x}_F - \myrandvector{x} = \mymatrix{K} \myrandvector{\tilde \mathtt{x}}_\cent, \\
		\mymatrix{\tilde P}_F &\eqdef \E\left[\myrandvector{\tilde x}_F \myrandvector{\tilde x}_F^{\intercal}\right] =  \mymatrix{K} \mymatrix{\tilde P}_\cent \mymatrix{K}^{\intercal},\label{eq: MSE original}
	\end{align}
\end{subequations}
where $\myrandvector{\tilde \mathtt{x}}_\cent \eqdef \begin{pmatrix} \myrandvector{\tilde x}_1^{\intercal} & \cdots & \myrandvector{\tilde x}_N^{\intercal} \end{pmatrix}^{\intercal} \in \R^{Nd}$ is the centralized vector of estimation errors, and $\mymatrix{\tilde P}_\cent \eqdef \E\left[\myrandvector{\tilde \mathtt{x}}_\cent \myrandvector{\tilde \mathtt{x}}_\cent^{\intercal}\right]$ is its covariance.
The objective of the optimal fusion is to minimize the estimation error, \ie{} to minimize some cost function on $\mymatrix{\tilde P}_F$, \eg{} its trace or its determinant.

If the covariance matrix $\mymatrix{\tilde P}_\cent$ is known, the optimal fusion is well known. It is recalled in the following lemma.
\begin{lem}[Optimal linear fusion]\label{lem: Optimal linear fusion}
	Let $\mymatrix{\tilde P}_\cent \succ \mymatrix{0}$ be the centralized covariance of the estimation errors, and define:
	\begin{equation}
		\mymatrix{K}^* = (\mymatrix{H}^{\intercal} \mymatrix{\tilde P}_\cent^{-1} \mymatrix{H})^{-1} \mymatrix{H}^{\intercal} \mymatrix{\tilde P}_\cent^{-1}.
	\end{equation}
	Then, for all gains $\mymatrix{K}$ satisfying \eqref{eq: Constraint unbiasedness}, $\mymatrix{\tilde P}_F(\mymatrix{K}) \succeq \mymatrix{\tilde P}_F(\mymatrix{K}^*)$.
\end{lem}
\begin{proof} \com{Provided in Appendix~\ref{proof: lem: Optimal linear fusion}.}
\end{proof}
In Lemma~\ref{lem: Optimal linear fusion}, the assumption $\mymatrix{\tilde P}_\cent \succ \mymatrix{0}$ is not restrictive. Indeed, if the covariance $\mymatrix{\tilde P}_\cent$ is singular either one component of $\myvector{x}$ is perfectly estimated by one of the $\myvector{\hat x}_i$ and it can be removed from the problem, or some components of the $\myvector{\tilde x}_i$ are perfectly correlated and considering the Moore–Penrose pseudo-inverse of $\mymatrix{\tilde P}_\cent$ solves the problem, see \eg{} \cite{li2003optimal}.

According to Lemma~\ref{lem: Optimal linear fusion}, $\mymatrix{\tilde P}_F^* \eqdef \mymatrix{\tilde P}_F(\mymatrix{K}^*) = (\mymatrix{H}^{\intercal} \mymatrix{\tilde P}_\cent^{-1} \mymatrix{H})^{-1}$ is the minimal covariance in the Loewner ordering sense. Therefore, no other gain can provide a better precision in any direction, and $\mymatrix{\tilde P}_F^*$ reaches the minimum for all increasing cost functions.
Lemma~\ref{lem: Optimal linear fusion} has a well-known special case. If the estimation errors are independent, \ie{} $\mymatrix{\tilde P}_{i,j} = \mymatrix{0}$ for all $i \ne j$. In this case, the optimal fusion is provided by the information filter, see \eg{} \cite{anderson1979optimal}:
\begin{subequations}
	\begin{align}
		\myrandvector{\hat x}_F(\mymatrix{K}^*) &= \mymatrix{\tilde P}_F^* \sum_{i = 1}^N \mymatrix{\tilde P}_i^{-1}\myrandvector{\hat x}_i, \\
		\mymatrix{\tilde P}_F^* &= \left(\sum_{i = 1}^N \mymatrix{\tilde P}_i^{-1}\right)^{-1}.
	\end{align}
\end{subequations}

\subsection{Conservative linear fusion}

Applying the optimal fusion of Lemma~\ref{lem: Optimal linear fusion} requires knowledge of the centralized matrix $\mymatrix{\tilde P}_\cent$. If the covariance matrix $\mymatrix{\tilde P}_\cent$ is (partially) unknown, \ie{} if some covariances $\mymatrix{\tilde P}_i$ or cross-covariances $\mymatrix{\tilde P}_{i,j}$ are unknown, the covariance matrix after fusion $\mymatrix{\tilde P}_F = \mymatrix{K} \mymatrix{\tilde P}_\cent \mymatrix{K}^{\intercal}$ cannot be computed, and \emph{a fortiori}, the optimal fusion cannot be implemented. In this case, an alternative is to provide a \emph{conservative} upper bound.

Generally, even if $\mymatrix{\tilde P}_\cent$ is unknown, it belongs to a given set $\rA$ of \emph{admissible} covariances.
For example, if the covariances $\mymatrix{\tilde P}_i$ are known but the cross-covariances $\mymatrix{\tilde P}_{i,j}$ are not, the set of admissible covariance is:
\begin{equation}\label{eq: Admissible set CI}
	\rA_{\mathrm{CI}} \eqdef \set{\mymatrix{P}_\cent \succeq \mymatrix{0} \suchthat \forall i, \ \mymatrix{P}_{i,i} = \mymatrix{\tilde P}_{i} 
	}.
\end{equation}
Note that $\mymatrix{\tilde P}_\cent$ (with a tilde) denotes the true but unknown covariance, while $\mymatrix{P}_\cent$ (without a tilde) is a generic notation for a covariance matrix in $\rA$. Conservative upper bounds are then defined as follows.

\begin{definition}[Conservative upper bound \cite{forsling2022conservative}]
	A matrix $\mymatrix{B}_F \succeq \mymatrix{0}$ is said to be a \emph{conservative upper bound} for the fusion induced by the gain $\mymatrix{K}$ over the set $\rA$, if:
	\begin{align*}
		\forall \mymatrix{P}_\cent &\in \rA, & 
		\mymatrix{\tilde P}_F(\mymatrix{K}, \mymatrix{P}_{\cent}) &\preceq \mymatrix{B}_F,
	\end{align*}
	with $\mymatrix{\tilde P}_F(\mymatrix{K}, \mymatrix{P}_{\cent}) \eqdef \mymatrix{K}\mymatrix{P}_{\cent} \mymatrix{K}^{\intercal}$.
\end{definition}

Considering a conservative bound $\mymatrix{B}_F$ ensures that the estimation error is not underestimated. In the algorithms, the uncomputable covariance $\mymatrix{\tilde P}_F(\mymatrix{K}, \mymatrix{\tilde P}_{\cent})$ is replaced by the overpessimistic but computable bound $\mymatrix{B}_F$. As the bound provides guarantees on the estimation error, it is sought to be as small as possible. The covariances are compared using a cost function $J$, typically the trace or the determinant. Finding the best conservative fusion consists in solving the following problem.
\begin{prob}[Optimal Conservative Linear Fusion]\label{pro: Optimal conservative fusion problem}
	\begin{equation}\label{eq: Optimal fusion problem}
		\left\{\begin{array}{cl}
			\minimize\limits_{\mymatrix{K}, \mymatrix{B}_F} & J(\mymatrix{B}_F) \\
			\subject{} & \mymatrix{K}\mymatrix{H} = \mymatrix{I}_d, \\
			& \forall \mymatrix{P}_{\cent} \in \rA, \, \mymatrix{\tilde P}_F(\mymatrix{K}, \mymatrix{P}_{\cent}) \preceq \mymatrix{B}_F.
		\end{array}\right.
	\end{equation}
\end{prob}
Problem~\ref{pro: Optimal conservative fusion problem} is recurrent in the literature, \eg{} different variants are discussed in \cite{reinhardt2015minimum ,ajgl2018fusion,forsling2022conservative}. The number of unknowns in \eqref{eq: Optimal fusion problem} is $O(d^2N^2)$. It can be numerically solved using robust semi-definite programming \cite{forsling2022conservative} but with a computation cost too high for real time applications. An important difference with the usual case where $\mymatrix{\tilde P}_\cent$ is known is that the solution of Problem~\ref{pro: Optimal conservative fusion problem} generally depends on the cost function $J$, \eg{} minimizing the trace or the determinant can result in different bounds. There is no minimum in the Loewner ordering sense.

\subsection{Covariance Intersection and Split Covariance Intersection}

This section recalls the derivation of CI and SCI and some important theoretical results on these two conservative fusion rules.

CI and SCI have both been proposed by Julier and Uhlmann \cite{uhlmann1996general, julier1997nondivergent, julier2001general}. They address two different admissible sets but are designed similarly. First an upper bound is provided for the set of admissible centralized covariances, then the bound on the fused estimator is constructed using the following lemma.
\begin{lem}\label{lem: Centralized bound to fused bound}
	Let $\rA$ be a set of admissible covariances and $\mymatrix{B}_\cent \succ \mymatrix{0}$. If for all $\mymatrix{P}_\cent \in \rA$, $\mymatrix{P}_\cent \preceq \mymatrix{B}_\cent$, then $\mymatrix{B}_F = (\mymatrix{H}^{\intercal} \mymatrix{B}_\cent^{-1} \mymatrix{H})^{-1}$ is a conservative bound over $\rA$ for the fusion induced by the gain $\mymatrix{K} = \mymatrix{B}_F \mymatrix{H}^{\intercal} \mymatrix{B}_\cent^{-1}$.
\end{lem}
\begin{proof} \com{Provided in Appendix~\ref{proof: lem: Centralized bound to fused bound}.}
\end{proof}

CI addresses the case where the covariances of the estimation errors,  $\mymatrix{\tilde P}_i$, are known, but not their cross-covariances, $\mymatrix{\tilde P}_{i,j}$. The set of admissible covariances considered by CI is therefore $\rA_{\mathrm{CI}}$ given in \eqref{eq: Admissible set CI}.
SCI is an extension of CI in which the estimation errors are assumed to be split into two components: $\myvector{\tilde x}_i = \myvector{\tilde x}_i^{(1)} + \myvector{\tilde x}_i^{(2)}$. The centralized vectors associated with the two components are denoted $\myrandvector{\tilde \mathtt{x}}_\cent^{(l)} \eqdef \begin{pmatrix} \myrandvector{\tilde x}_1^{(l)\intercal} & \cdots & \myrandvector{\tilde x}_N^{(l)\intercal}\end{pmatrix}^{\intercal}$, $l \in \set{1, 2}$. The covariances of the $2N$ components are known and denoted as $\mymatrix{\tilde P}_i^{(l)} \eqdef \E\left[\myrandvector{\tilde x}_i^{(l)} \myrandvector{\tilde x}_i^{(l)\intercal}\right]$. The cross-covariances between the first components, $\mymatrix{\tilde P}^{(1)}_{i,j}$, are unknown, while the second components, $\myrandvector{\tilde x}_i^{(2)}$, are uncorrelated with each other and with the $\myrandvector{\tilde x}_i^{(1)}$. The set of admissible covariance considered by SCI is therefore:
\begin{equation}\label{eq: Admissible set SCI}
	\rA_{\mathrm{SCI}} \eqdef \set{\mymatrix{P}^{(1)}_\cent + \mymatrix{\tilde P}^{(2)}_\cent \suchthat \forall i, \ \mymatrix{P}^{(1)}_{i,i} = \mymatrix{\tilde P}^{(1)}_{i}, \mymatrix{P}^{(1)}_\cent \succeq \mymatrix{0} 
	}.
\end{equation}
In \eqref{eq: Admissible set SCI}, the matrix $\mymatrix{\tilde P}_\cent^{(2)} \eqdef \E\left[\myrandvector{\tilde \mathtt{x}}_\cent^{(2)} \myrandvector{\tilde \mathtt{x}}_\cent^{(2)\intercal}\right]$ is known, $\mymatrix{\tilde P}_\cent^{(2)} =\diag(\mymatrix{\tilde P}_1^{(2)}, \dots, \mymatrix{\tilde P}_N^{(2)})$, while $\mymatrix{P}^{(1)}_\cent$ represents the unknown covariance $\E\left[\myrandvector{\tilde \mathtt{x}}_\cent^{(1)} \myrandvector{\tilde \mathtt{x}}_\cent^{(1)\intercal}\right]$. CI corresponds to the particular case of SCI without uncorrelated components, \ie{} $\mymatrix{\tilde P}_\cent^{(2)} = \mymatrix{0}$.

Recall, that $\rK^N$ denotes the unit simplex of $\R^N$, CI and SCI rely on the following family of centralized bounds.

\begin{lem}[CI centralized bound \cite{julier1997nondivergent}]\label{lem: CI centralized bound}
	Let $\myvector{\omega} \in \interior \rK^N$, and define:
	\begin{equation}
		\mymatrix{B}_\cent^{\mathrm{CI}}(\myvector{\omega}) \eqdef \diag\left(\frac{1}{\omega_1} \mymatrix{\tilde P}_1, \dots, \frac{1}{\omega_N} \mymatrix{\tilde P}_N\right).
	\end{equation}
	Then, for all $\mymatrix{P}_\cent \in \rA_{\mathrm{CI}}$, $\mymatrix{P}_\cent \preceq \mymatrix{B}_\cent^{\mathrm{CI}}(\myvector{\omega})$.
\end{lem}
\begin{proof} \com{Provided in Appendix~\ref{proof: lem: CI centralized bound}.}
\end{proof}
Using Lemma~\ref{lem: CI centralized bound}, it is straightforward to show that for all $\mymatrix{P}_\cent \in \rA_{\mathrm{SCI}}$ and all $\myvector{\omega} \in \interior \rK^N$: $\mymatrix{P}_\cent \preceq \mymatrix{B}_\cent^{(1)}(\myvector{\omega}) + \mymatrix{\tilde P}_{\cent}^{(2)}$ where:
\begin{equation}
	\mymatrix{B}_\cent^{(1)}(\myvector{\omega}) \eqdef \diag\left(\frac{1}{\omega_1} \mymatrix{\tilde P}_1^{(1)}, \dots, \frac{1}{\omega_N} \mymatrix{\tilde P}_N^{(1)}\right).
\end{equation}

The CI and SCI fusion rules are then deduced from Lemma~\ref{lem: Centralized bound to fused bound}.
\begin{definition}[CI fusion rule \cite{julier1997nondivergent}]
	For all $\myvector{\omega} \in \rK^N$, the CI fused estimator is defined as:
	\begin{subequations}\label{eq: CI equations}
		\begin{align}
			\myrandvector{\hat x}_F^{\mathrm{CI}}(\myvector{\omega}) &\eqdef \mymatrix{B}_F^{\mathrm{CI}}(\myvector{\omega}) \sum_i \omega_i \mymatrix{\tilde P}_i^{-1} \myrandvector{\hat x}_i, \\
			\mymatrix{B}_F^{\mathrm{CI}}(\myvector{\omega}) &\eqdef \left(\sum_i \omega_i \mymatrix{\tilde P}_i^{-1}\right)^{-1}.
		\end{align}
	\end{subequations}
\end{definition}
\begin{definition}[SCI fusion rule \cite{julier2001general}]
	For all $\myvector{\omega} \in \rK^N$, the SCI fused estimator is defined as:
	\begin{subequations}\label{eq: SCI equations}
		\begin{align}
			\myrandvector{\hat x}_F^{\mathrm{SCI}}(\myvector{\omega}) &\eqdef \mymatrix{B}_F^{\mathrm{SCI}}(\myvector{\omega}) \sum_i \omega_i (\mymatrix{\tilde P}_i^{(1)} + \omega_i \mymatrix{\tilde P}_i^{(2)})^{-1} \myrandvector{\hat x}_i, \\
			\mymatrix{B}_F^{\mathrm{SCI}}(\myvector{\omega}) &\eqdef \left[\sum_i \omega_i (\mymatrix{\tilde P}_i^{(1)} + \omega_i \mymatrix{\tilde P}_i^{(2)})^{-1}\right]^{-1}.
		\end{align}
	\end{subequations}
\end{definition}
CI and SCI equations are valid for all $\myvector{\omega} \in \rK^N$ by continuity. Having some $\omega_i = 0$ is equivalent to exclude the $i$th estimation. Examples of bounds generated by CI and SCI are illustrated in Figure~\ref{sfig: CI bounds} and Figure~\ref{sfig: SCI bounds}. The name Covariance Intersection comes from the fact that, the ellipsoids associated with CI bounds contain the intersection of the ellipsoids of the covariances $\mymatrix{\tilde P}_i$. In both schemes, the parameter $\myvector{\omega}$ must be chosen. It can be optimized to minimize the cost function $J$, such an optimization is now only on $N-1$ variables.
As $N$ increases, the optimization of the weights becomes harder and sub-optimal choices have been proposed to speed up the implementation \cite{niehsen2002information, franken2005improved, deng2012sequential}.

CI and SCI have been applied to a wide range of problems: distributed estimation \cite{arambel2001covariance, hu2011diffusion}, simultaneous localization and mapping (SLAM) \cite{julier2007using}, cooperative positioning \cite{li2013split, pierre2018range, lai2019cooperative}, image processing \cite{guo2010covariance}, or the monitoring of vital signs \cite{zhang2020covariance}. The splitting assumption on the estimation errors appears naturally when dealing with distributed and cooperative systems as detailed in the Section~\ref{sec: Application}.
Furthermore, the good performances of CI and SCI are justified by theoretical results. CI was first proved to provide the optimal bound with respect to the trace for the fusion of $N = 2$ estimators \cite{chen2002estimation}. Then, this optimality was extended to any increasing cost function $J$ in \cite{reinhardt2015minimum}. For SCI, only special cases have been discussed. For example, in \cite{wu2017covariance}, the authors study the fusion of $N = 2$ estimators and consider the case where the Pearson correlation coefficient is bounded by a given value $\rho \in [0,1]$. That case corresponds to considering the set $\rA_{\mathrm{SCI}}$ with $\mymatrix{\tilde P}_i^{(1)} = \rho \mymatrix{\tilde P}_i$ and $\mymatrix{\tilde P}_i^{(2)} = (1-\rho) \mymatrix{\tilde P}_i$. The authors proved that in that case SCI, provide the trace optimal fusion bound.

In Section~\ref{sec: Optimality}, the optimality of SCI for the fusion of two estimators will be generalized to any increasing cost function and a wider class of context. Before that, the next section introduces an extension of SCI motivated by the distributed estimation problems and the will of exploiting the correlated components during the fusion.

\section{Extended Split Covariance Intersection}\label{sec: ESCI}

\subsection{Motivating example}\label{ssec: Motivating example}

To motivate the need for a new fusion rule, consider the following example inspired by the problem of distributed estimation. This problem will be solved in Section~\ref{sec: Application}. In this example, only one iteration of the distributed estimation algorithm is considered.

Consider a random dynamic state $\myrandvector{x}$ observed by a network of $N$ sensors. The dynamics of $\myrandvector{x}$ and the measurements of a Node~$i \in \intEnt{1}{N}$ are modeled as:
\begin{subequations}\label{eq: Motivation model}
	\begin{align}
		\myrandvector{x} &= \mymatrix{F}\myrandvector{x}^- + \myrandvector{w}, \\
		\myrandvector{z}_i &= \mymatrix{H}_i \myrandvector{x} + \myrandvector{v}_i. \label{eq: Example spliting mesure SCI}
	\end{align}
\end{subequations}
In \eqref{eq: Motivation model}, $\mymatrix{F}$ is the evolution matrix, $\mymatrix{H}_i$ is the observation matrix of Node~$i$, $\myrandvector{x}^-$ is the previous state, $\myrandvector{w}$ is the process noise, and $\myrandvector{v}_i$ is the measurement noise of Node~$i$. The process noise and the measurement noise are assumed to be zero-mean, with known covariances, and independent of each other.

Suppose each Node~$i$ has an unbiased estimator of $\myrandvector{x}^-$ denoted as $\myrandvector{\hat x}_i^{-}$. \com{This unbiased estimator was generated during the previous iterations (see Section~\ref{sec: Application} for an application).} Then, Node~$i$ produces a new estimator of $\myrandvector{x}$ in a Kalman filter fashion:
\begin{equation}\label{eq: Example spliting update SCI}
	\myrandvector{\hat x}_i = (\mymatrix{I} - \mymatrix{W}_i \mymatrix{H}_i)\mymatrix{F} \myrandvector{\hat x}_i^{-} + \mymatrix{W}_i \myrandvector{z}_i,
\end{equation}
where $\myvector{W}_i$ is the Kalman gain.
The new estimates are shared through the network, and Node~$i$ fuses its estimate with those received from its neighbors. The cross-covariances between the errors on the $\myrandvector{\hat x}_i^{-}$ depend on the fusions performed by all the agents on the previous iterations. They are hardly tractable because of time and communication constraints. These cross-covariances are therefore assumed to be unknown, and the fusion of the resulting $\myrandvector{\hat x}_i$ must be performed conservatively. Even if the correlations between the $\myrandvector{\tilde x}_i^-$ are unknown, the errors $\myrandvector{\tilde x}_i$ do have a structure. Based on \eqref{eq: Motivation model} and \eqref{eq: Example spliting update SCI}:
\begin{equation}\label{eq: Example SCI decomposition}
	\myrandvector{\tilde x}_i = \underbrace{(\mymatrix{I} - \mymatrix{W}_i \mymatrix{H}_i)\mymatrix{F} \myrandvector{\tilde x}_i^{-} - (\mymatrix{I} - \mymatrix{W}_i \mymatrix{H}_i)\myrandvector{w}}_{\myrandvector{\tilde x}_i^{(1)}} +\underbrace{ \mymatrix{W}_i \myrandvector{v}_i}_{\myrandvector{\tilde x}_i^{(2)}},
\end{equation}
where $\myvector{\tilde x}_i^{(2)}$ denotes the independent term of the error. The fusion of the $\myrandvector{\hat x}_i$ can be performed using CI or SCI. With CI, the structure of the errors is not exploited. With SCI, the independent components induced by the measurement noises can be isolated and exploited to produce smaller bounds.
However, in the SCI decomposition \eqref{eq: Example SCI decomposition}, the common process noise is not exploited. SCI can only isolate uncorrelated component, so it is not suited to treat such common terms.
The extension of  SCI proposed in the next section is designed to take into account these common terms. With the new ESCI fusion rule, all the terms with known correlations are gathered in $\myrandvector{\tilde x}_i^{(2)}$. The splitting of the errors will be:
\begin{equation}\label{eq: Example ESCI decompostion}
	\myrandvector{\tilde x}_i = \underbrace{(\mymatrix{I} - \mymatrix{W}_i \mymatrix{H}_i)\mymatrix{F} \myrandvector{\tilde x}_i^{-}}_{\myrandvector{\tilde x}_i^{(1)}} \underbrace{- (\mymatrix{I} - \mymatrix{W}_i \mymatrix{H}_i)\myrandvector{w} + \mymatrix{W}_i \myrandvector{v}_i}_{\myrandvector{\tilde x}_i^{(2)}}.
\end{equation}
This new decomposition allows to isolate in the first components only the terms whose correlations are unknown, \ie{} which may be correlated to any degree. That was not the case with the decomposition \eqref{eq: Example SCI decomposition}: for example, the first components in \eqref{eq: Example SCI decomposition} cannot be perfectly negatively correlated.

Figure~\ref{fig: Comparison of the bounds} presents the fusion bounds obtained with CI, SCI, and the new ESCI (described in the next section). To produce this figure, the evolution matrix was set to $\mymatrix{F} = \mymatrix{I}_2$, and the observation matrices to $\mymatrix{H}_1 = \begin{bmatrix} 1 & 0\end{bmatrix}$ and $\mymatrix{H}_2 = \begin{bmatrix} 0 & 1\end{bmatrix}$. The covariance of the process noise $\myrandvector{w}$ was set to $\mymatrix{Q} = 4 \mymatrix{I}_2$, and the covariances of the measurement noises $\myrandvector{v}_1$ and $\myrandvector{v}_2$ to $\mymatrix{R}_1 = \mymatrix{R}_2 = 9$. Finally, the covariances of the initial estimation errors were set to:
\begin{align*}
	\mymatrix{\tilde P}_1^{-} &= \begin{bmatrix} 1 & -1 \\ -1 & 4\end{bmatrix}, &
	\mymatrix{\tilde P}_2^{-} &= \begin{bmatrix} 8 & 3 \\ 3 & 2\end{bmatrix}.
\end{align*}
On that toy example, it can be observed that the ESCI bounds are tighter than the SCI (and than the CI) bounds. Thus, considering the process noise allows to get better error guarantees.

The next section introduces the new ESCI fusion rule.

\subsection{ESCI fusion rule}

Consider a random state $\myrandvector{x} \in \R^d$ and $N \ge 2$ unbiased estimators of $\myrandvector{x}$ denoted as $\myrandvector{\hat x}_i$ for $i \in \intEnt{1}{N}$. As for SCI, the estimation errors are assumed to be split into two components $\myrandvector{\tilde x}_i = \myrandvector{\tilde x}_i^{(1)} + \myrandvector{\tilde x}_i^{(2)}$, and the covariances of the $2N$ components are known and denoted as $\mymatrix{\tilde P}_i^{(l)} \eqdef \E\left[\myrandvector{\tilde x}_i^{(l)} \myrandvector{\tilde x}_i^{(l)\intercal}\right]$, $l\in\set{1,2}$. The first components are still correlated to an unknown degree, \ie{} the cross-covariances $\mymatrix{\tilde P}_{i,j}^{(1)} \eqdef \E\left[\myrandvector{\tilde x}_i^{(1)} \myrandvector{\tilde x}_j^{(1)\intercal}\right]$ are unknown. Unlike SCI, the second components are not assumed to be uncorrelated, but to have known second-order moments: the covariances $\mymatrix{\tilde P}_\cent^{(2)}\eqdef \E\left[\myrandvector{\tilde \mathtt{x}}_\cent^{(2)} \myrandvector{\tilde \mathtt{x}}_\cent^{(2)\intercal}\right]$ and $\mymatrix{\tilde P}_\cent^{(1,2)} \eqdef \E\left[\myrandvector{\tilde \mathtt{x}}_\cent^{(1)} \myrandvector{\tilde \mathtt{x}}_\cent^{(2)\intercal}\right]$ are known.

\com{In the following, it is assumed without loss of generality that $\mymatrix{\tilde P}_\cent^{(1,2)} = \mymatrix{0}$. Indeed, if the cross-covariance $\mymatrix{\tilde P}_\cent^{(1,2)} \ne \mymatrix{0}$, the errors can be made uncorrelated by applying the following transformation:}
\begin{subequations}
	\begin{align}
		\myrandvector{\tilde \mathtt{x}}_\cent^{(1)\prime} &\gets \myrandvector{\tilde \mathtt{x}}_\cent^{(1)} - \mymatrix{\tilde P}_{\cent}^{(1,2)}(\mymatrix{\tilde P}^{(2)}_\cent)^{-1}\myrandvector{\tilde \mathtt{x}}_\cent^{(2)} \\
		\myrandvector{\tilde \mathtt{x}}_\cent^{(2)\prime} &\gets \myrandvector{\tilde \mathtt{x}}_\cent^{(2)} + \mymatrix{\tilde P}_{\cent}^{(1,2)}(\mymatrix{\tilde P}^{(2)}_\cent)^{-1}\myrandvector{\tilde \mathtt{x}}_\cent^{(2)}
	\end{align}
\end{subequations}
After this \com{transformation}, the errors $\myrandvector{\tilde \mathtt{x}}_\cent^{(1)\prime}$ and $\myrandvector{\tilde \mathtt{x}}_\cent^{(2)\prime}$ satisfy the same hypothesis, only the cross-covariances between the $\myrandvector{\tilde x}_{i}^{(1)\prime}$ are unknown, but $\E\left[\myrandvector{\tilde \mathtt{x}}_\cent^{(1)\prime}\myrandvector{\tilde \mathtt{x}}_\cent^{(2)\prime\intercal}\right] = \mymatrix{0}$. With this additional assumptions the set of admissible covariances is:
\begin{equation}\label{eq: Admissible set ESCI}
	\rA_{\mathrm{ESCI}} \eqdef \set{\mymatrix{P}^{(1)}_\cent + \mymatrix{\tilde P}^{(2)}_\cent \suchthat \forall i, \ \mymatrix{P}^{(1)}_{i,i} = \mymatrix{\tilde P}^{(1)}_{i}, \mymatrix{P}^{(1)}_\cent \succeq \mymatrix{0} 
	}
\end{equation}
The sets $\rA_{\mathrm{ESCI}}$ and $\rA_{\mathrm{SCI}}$ \eqref{eq: Admissible set SCI} have the same expression. The only difference is that, in \eqref{eq: Admissible set ESCI}, the covariance $\mymatrix{\tilde P}^{(2)}_\cent$ is not necessary a block-diagonal matrix.

The ESCI fusion rule is constructed similarly as the SCI fusion rule. First, the following family of upper bounds is defined.
\begin{subequations}
	\begin{align}
		\mymatrix{B}_\cent^{\mathrm{ESCI}}(\myvector{\omega}) &\eqdef \mymatrix{B}_\cent^{(1)}(\myvector{\omega}) + \mymatrix{\tilde P}_\cent^{(2)}, \label{eq: Centralized bound ESCI} \\
		\mymatrix{B}_\cent^{(1)}(\myvector{\omega}) &\eqdef \diag\left(\frac{1}{\omega_1} \mymatrix{\tilde P}_1^{(1)}, \dots, \frac{1}{\omega_N} \mymatrix{\tilde P}_N^{(1)}\right).
	\end{align}
\end{subequations}
From Lemma~\ref{lem: CI centralized bound}, for all $\mymatrix{P}_\cent \in \rA_{\mathrm{ESCI}}$ and all $\myvector{\omega} \in \interior \rK^N$, $\mymatrix{P}_\cent \preceq \mymatrix{B}_\cent^{\mathrm{ESCI}}(\myvector{\omega})$.
The new fusion rule is then defined by applying Lemma~\ref{lem: Centralized bound to fused bound}.
\begin{definition}[ESCI fusion rule]
	For all $\myvector{\omega} \in \rK^N$, the ESCI fused estimator is defined as:
	\begin{subequations}\label{eq: ESCI equations}
		\begin{align}
			\myrandvector{\hat x}_F^{\text{ESCI}}(\myvector{\omega}) &\eqdef \mymatrix{B}^{\mathrm{ESCI}}_F(\myvector{\omega})\mymatrix{H}^{\intercal} \mymatrix{B}_\cent^{\mathrm{ESCI}}(\myvector{\omega})^{-1} \myrandvector{\hat \mathtt{x}}_\cent \\
			\mymatrix{B}^{\mathrm{ESCI}}_F(\myvector{\omega}) &\eqdef \left(\mymatrix{H}^{\intercal} \mymatrix{B}_\cent^{\mathrm{ESCI}}(\myvector{\omega})^{-1} \mymatrix{H}\right)^{-1}
		\end{align}
	\end{subequations}
	It is conservative over the set of admissible covariance $\rA_{\mathrm{ESCI}}$.
\end{definition}

The ESCI fusion rule extends several well-known fusions. First, it is an extension of the SCI fusion rule. If the second components are uncorrelated, \ie{} if $\mymatrix{\tilde P}_{i,j}^{(2)} = \mymatrix{0}$, then \eqref{eq: SCI equations} and \eqref{eq: ESCI equations} coincide. Unfortunately, if the matrix $\mymatrix{\tilde P}_\cent^{(2)}$ has no structure, \eqref{eq: ESCI equations} cannot be simplified.

As an extension of SCI, ESCI is also a generalization of CI. CI and ESCI coincide when \com{there are no known components}, \ie{} when $\mymatrix{\tilde P}_\cent^{(2)} = \mymatrix{0}$. Conversely, if there are no unknown components, \ie{} if all $\myrandvector{\tilde x}_i^{(1)} = \myvector{0}$, then the centralized bound does not depend on $\myvector{\omega}$, since $\mymatrix{B}_\cent^{(1)} = \mymatrix{0}$. In this case, the ESCI fusion coincides as expected with the optimal bound provided in Lemma~\ref{lem: Optimal linear fusion}.
Furthermore, the ESCI fusion rule is also equivalent to the PCI fusion rule \cite{ajgl2019rectification} when the errors are partitioned. \com{The term \guillemets{partitioned} means that the estimates, and therefore their errors, are written as $\myrandvector{\hat x}_i = \begin{pmatrix}	\myrandvector{\hat x}_i^{(a)\intercal} & \myrandvector{\hat x}_i^{(b)\intercal} \end{pmatrix}^{\intercal}$. For example, in tracking applications, $\myrandvector{\hat x}_i^{(a)}$ may stand for the position of a target and $\myrandvector{\hat x}_i^{(b)}$ for its velocity. In PCI, only the cross-covariances between the errors $\myrandvector{\tilde x}_i^{(a)}$ are unknown which corresponds to considering the splitting:}
\begin{align*}
	\myrandvector{\tilde x}_i^{(1)} &= \begin{pmatrix} \myrandvector{\tilde x}_i^{(a)\intercal} & \myvector{0}^{\intercal} \end{pmatrix}^{\intercal}, &
	\myrandvector{\tilde x}_i^{(2)} &= \begin{pmatrix} \myvector{0}^{\intercal} & \myrandvector{\tilde x}_i^{(b)\intercal} \end{pmatrix}^{\intercal}.
\end{align*}
\com{A more detailed discussion on the relations between ESCI, PCI and SCI can be found in \cite{ajgl2022linear}.}

\subsection{Particular case of a common noise}\label{ssec: Particular case of a common noise}

In many applications, including distributed estimation, the correlated terms in the second components are induced by a common noise. In this case, the computation of the ESCI fusion rule can be simplified. Consider that the estimation errors are decomposed as:
\begin{equation}\label{eq: Error decomposition with common noise}
	\myrandvector{\tilde x}_i = \myrandvector{\tilde x}_i^{(1)} + \myrandvector{\tilde x}_i^{(\mathrm{ind})} + \mymatrix{M}_i \myrandvector{w},
\end{equation}
where $\myrandvector{w}$ is a zero-mean random noise, the $\mymatrix{M}_i$ are known matrices, and the components $\myrandvector{\tilde x}_i^{(\mathrm{ind})}$ are zero-mean and uncorrelated between each other, with the $\myrandvector{\tilde x}_i^{(1)}$ and with the noise $\myrandvector{w}$. The covariances of each components are known and denoted as $\mymatrix{Q}$ for the noise $\myrandvector{w}$ and $\mymatrix{\tilde P}_i^{(\mathrm{ind})}$ for the $\myrandvector{\tilde x}_i^{(\mathrm{ind})}$. The decomposition \eqref{eq: Error decomposition with common noise} applies in particular to the motivating example introduced in Section~\ref{ssec: Motivating example}:
\begin{equation}
	\myrandvector{\tilde x}_i = \underbrace{(\mymatrix{I} - \mymatrix{W}_i \mymatrix{H}_i)\mymatrix{F} \myrandvector{\tilde x}_i^{-}}_{\myrandvector{\tilde x}_i^{(1)}} - \underbrace{ (\mymatrix{I} - \mymatrix{W}_i \mymatrix{H}_i)}_{\mymatrix{M}_i}\myrandvector{w} + \underbrace{\mymatrix{W}_i \myrandvector{v}_i}_{\myvector{\tilde x}_i^{(\mathrm{ind})}}.
\end{equation}

With such a decomposition, the centralized bound \eqref{eq: Centralized bound ESCI} becomes:
\begin{equation}
	\mymatrix{B}_\cent^{\mathrm{ESCI}}(\myvector{\omega}) = \mymatrix{B}_\cent^{(1)}(\myvector{\omega}) + \mymatrix{\tilde P}_\cent^{(\mathrm{ind})} + \mymatrix{M}_\cent \mymatrix{Q} \mymatrix{M}_\cent^{\intercal},
\end{equation}
with $ \mymatrix{\tilde P}_\cent^{(\mathrm{ind})} \eqdef \diag\left(\mymatrix{\tilde P}_1^{(\mathrm{ind})},\, \dots,  \mymatrix{\tilde P}_N^{(\mathrm{ind})}\right)$ and $\mymatrix{M}_\cent \eqdef \begin{bmatrix}\mymatrix{M}_1^{\intercal} & \cdots & \mymatrix{M}_N^{\intercal} \end{bmatrix}^{\intercal}$. By noticing that $\mymatrix{B}_\cent^{(1)}(\myvector{\omega}) + \mymatrix{\tilde P}_\cent^{(\mathrm{ind})}$ is a block diagonal matrix and using the Woodbury inversion identity the ESCI fusion rule becomes:
\begin{subequations}\label{eq: Common noise ESCI equations}
\begin{align}
	\myrandvector{\hat x}_F &= \mymatrix{B}_F \sum_i \omega_i \left(\mymatrix{I} - \mymatrix{S}_1 \mymatrix{S}_0^{-1} \mymatrix{M}_i^{\intercal}\right)\mymatrix{\tilde P}_i^{\prime-1}\myrandvector{\hat x}_i, \\
	\mymatrix{B}_F &= \left(\sum_i \omega_i \mymatrix{\tilde P}_i^{\prime-1} - \mymatrix{S}_1 \mymatrix{S_0}^{-1} \mymatrix{S}_1^{\intercal} \right)^{-1},
\end{align}
where $\mymatrix{\tilde P}_i^{\prime} \eqdef \mymatrix{\tilde P}_i^{(1)} + \omega_i \mymatrix{\tilde P}_i^{(\mathrm{ind})}$ and:
\begin{align}
	\mymatrix{S}_0 &\eqdef \sum_i \omega_i \mymatrix{M}_i^{\intercal} \mymatrix{\tilde P}_i^{\prime-1} \mymatrix{M}_i + \mymatrix{Q}^{-1}, \\
	\mymatrix{S}_1 &\eqdef  \sum_i \omega_i \mymatrix{\tilde P}_i^{\prime-1} \mymatrix{M}_i.
\end{align}
\end{subequations}

The advantage of this new fusion rule over \eqref{eq: ESCI equations} is the computation cost. Equations \eqref{eq: ESCI equations} requires to invert the centralized covariance matrix which has size $Nd$, whereas \eqref{eq: Common noise ESCI equations} requires to invert $N+2$ covariance matrices of size $d$. As the computation cost of the inversion of matrix of size $n$ is  $O(n^3)$, the computation cost of the standard ESCI fusion rule \eqref{eq: ESCI equations} is $O(N^3 d^3)$ while the computation cost of \eqref{eq: Common noise ESCI equations} is $O(N d^3)$. Using \eqref{eq: Common noise ESCI equations} is therefore more efficient for large $N$.

\section{Optimality of ESCI for the fusion of two estimators}\label{sec: Optimality}

\subsection{Main result}

This section presents a theoretical result. The ESCI fusion rule provides the optimal conservative bound over the set $\rA_{\mathrm{ESCI}}$ for the fusion of two estimators. Throughout this section, the number of estimators is set to $N = 2$, and $J$ denotes an increasing cost function. Let us recall the problem of optimal conservative fusion in the context of ESCI.

\begin{prob}[Optimal Fusion with Split Errors]\label{pro: Main problem}
	\begin{equation*}
		\left\{\begin{array}{cl}
			\arg\min\limits_{\mymatrix{K}, \mymatrix{B}_F} & J(\mymatrix{B}_F) \\
			\subject{} & \mymatrix{K}\mymatrix{H} = \mymatrix{I} \\
			& \forall \mymatrix{P}_\cent \in \rA_{\mathrm{ESCI}}, \,  \mymatrix{\tilde P}_F(\mymatrix{K}, \mymatrix{P}_{\cent}) \preceq \mymatrix{B}_F
		\end{array}\right.
	\end{equation*}
	with $\mymatrix{\tilde P}_F(\mymatrix{K}, \mymatrix{P}_{\cent}) \eqdef \mymatrix{K} \mymatrix{P}_{\cent}\mymatrix{K}^{\intercal}$.
\end{prob}

The main result is the following theorem.

\begin{thm}\label{the: Main result}
	Let $(\mymatrix{K}, \mymatrix{B}_F)$ define a conservative fusion. $(\mymatrix{K}, \mymatrix{B}_F)$ is a solution of Problem~\ref{pro: Main problem} if and only if there exists $\myvector{\omega}^* \in \arg\min_{\myvector{\omega} \in \rK^2} J(\mymatrix{B}_{F}^{\mathrm{ESCI}}(\myvector{\omega}))$ such that $\mymatrix{B}_F = \mymatrix{B}_{F}^{\mathrm{ESCI}}(\myvector{\omega}^*)$.
\end{thm}

In particular, ESCI gives a solution to Problem~\ref{pro: Main problem}.
\begin{cor}
	Let $\myvector{\omega}^* \in \arg\min_{\myvector{\omega }\in \rK^2} J(\mymatrix{B}_F^{\mathrm{ESCI}}(\myvector{\omega}))$.
	A solution of Problem~\ref{pro: Main problem} is:
	\begin{subequations}\begin{align}
			\mymatrix{K}^* &= \mymatrix{B}_F^{\mathrm{ESCI}}(\myvector{\omega}^*) \mymatrix{H}^{\intercal} [\mymatrix{B}_\cent^{\mathrm{ESCI}}(\myvector{\omega}^*)]^{-1} \\
			\mymatrix{B}_F^* &= \mymatrix{B}_F^{\mathrm{ESCI}}(\myvector{\omega}^*)
	\end{align}\end{subequations}
\end{cor}
The rest of this section proves Theorem~\ref{the: Main result}. The implications of these two results are discussed in Section~\ref{sec: Discussion}.

\subsection{Sketch of the proof and preliminaries}

The method for proving Theorem~\ref{the: Main result} is inspired by the one used in \cite{reinhardt2015minimum} to prove the optimality of CI. First in Section~\ref{ssec: Minimal volume}, a minimal volume that all conservative bounds must contain is introduced and characterized. Then in Section~\ref{ssec: Tightness}, it is proved that this minimal volume is \emph{tightly circumscribed} only by ESCI bounds. Finally, the proof of Theorem~\ref{the: Main result} is given in Section~\ref{ssec: Proof of the main result}.
The main difference with the work in \cite{reinhardt2015minimum} is that the minimal volume for ESCI is much more complex than for CI. Although the chain of arguments is conceptually the same, the extension of the result in \cite{reinhardt2015minimum} is not straightforward and requires a careful redevelopment of the arguments.

In order to lighten the notation and as $\rK^2 = \set{(\omega, 1-\omega), \ \omega \in [0,1]}$, the bounds will be reparameterized by $\omega \in [0,1]$. For example, the slightly abusive notation $\mymatrix{B}_F^{\mathrm{ESCI}}(\omega)$ is used. Furthermore, $\bar\omega$ will denote $\bar\omega \eqdef 1 - \omega$.

\com{Finally, for the sake of clarity, Theorem~\ref{the: Main result} is first proved in the special case of positive definite covariance matrices. Considering positive definite covariance matrices allows to avoid singular cases and simplifies the proof. Theorem~\ref{the: Main result} still holds for positive semi-definite matrices: the proof for the general case requires an additional discussion, which is given in Section~\ref{ssec: Proof of the main result}. Therefore, the following assumption is made throughout this section.}
\begin{assumption}\label{as: Positivity}
	The covariances $\mymatrix{\tilde P}_1^{(1)}$, $\mymatrix{\tilde P}_2^{(1)}$ and $\mymatrix{\tilde P}_\cent^{(2)}$ are positive definite.
\end{assumption}

\subsection{Minimal volume}\label{ssec: Minimal volume}

This section introduces a minimal volume that all the ellipsoids associated with conservative upper bounds must contain. The discussion presented in this section was introduced in \cite{reinhardt2015minimum} and is valid for any admissible set $\rA$. It is based on the following lemma.

\begin{lem}\label{lem: Inequality optimal bounds}
	Let $\rA$ be a set of admissible covariances and $\mymatrix{B}_F$ be a conservative bound over $\rA$ for the fusion induced by any fusion gain $\mymatrix{K}$ satisfying \eqref{eq: Constraint unbiasedness}. Then,
	\begin{align}
		\forall \mymatrix{P}_\cent &\in \rA, & \mymatrix{\tilde P}_F^*(\mymatrix{P}_\cent) \preceq \mymatrix{B}_F,
	\end{align}
	where $\mymatrix{\tilde P}_F^*(\mymatrix{P}_\cent) \eqdef (\mymatrix{H}^{\intercal} \mymatrix{P}_\cent^{-1} \mymatrix{H})^{-1}$ is the optimal covariance given in Lemma~\ref{lem: Optimal linear fusion}.
\end{lem}
\begin{proof} \com{Provided in Appendix~\ref{proof: lem: Inequality optimal bounds}.}
\end{proof}
Geometrically, Lemma~\ref{lem: Inequality optimal bounds} means that the ellipse associated with any conservative bound $\mymatrix{B}_F$ must contain the volume:
\begin{equation}
	\rV(\rA) \eqdef \bigcup \rE(\mymatrix{\tilde P}_F^*(\mymatrix{P}_\cent)).
\end{equation}
$\rV(\rA)$ is the minimal volume associated with the set $\rA$.

\begin{figure*}
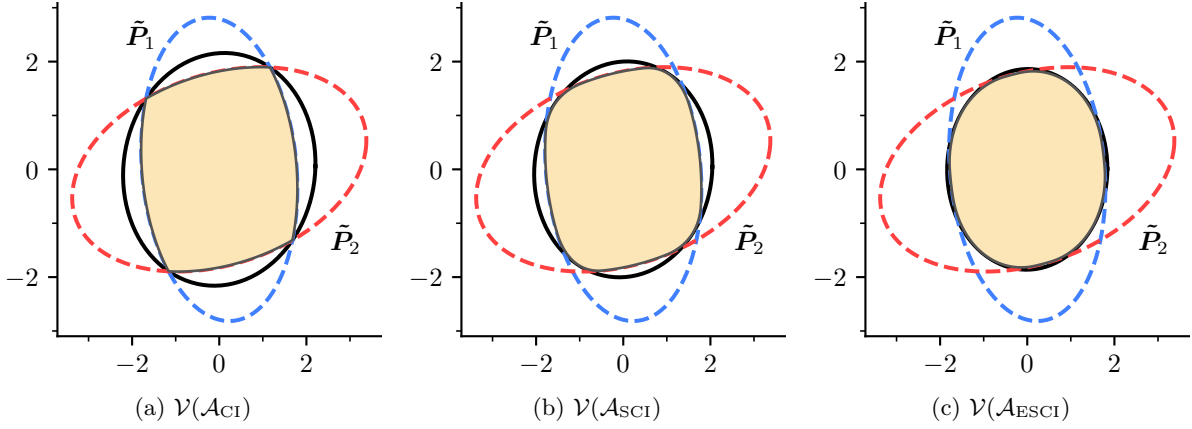

	\centering
	\null\hfill
	\subfloat[$\rV(\rA_{\mathrm{CI}})$]{\input{fig/volume_minimal_ic.pgf}}
	\hfill
	\subfloat[$\rV(\rA_{\mathrm{SCI}})$]{\input{fig/volume_minimal_icf.pgf}}
	\hfill
	\subfloat[$\rV(\rA_{\mathrm{ESCI}})$]{\input{fig/volume_minimal_icfg.pgf}}
	\hfill\null
	\caption{Comparison of the minimal volumes of CI, SCI and ESCI. The minimal volumes $\rV(\rA)$ are represented by the colored areas. The dark ellipses represent the bounds obtained with corresponding fusion for the parameter $\myvector{\omega} = (0.5,\ 0.5)$. The same configuration as in Fig.~\ref{fig: Comparison of the bounds} was used.}
	\label{fig: Minimal volumes}
\end{figure*}

The minimal volumes associated with the admissible sets $\rA_{\mathrm{CI}}$, $\rA_{\mathrm{SCI}}$ and $\rA_{\mathrm{ESCI}}$ are illustrated in Figure~\ref{fig: Minimal volumes}. In each case, a conservative bound has been plotted to illustrate that indeed, $\rV(\rA) \subseteq \rE(\mymatrix{B}_F)$.
For CI, the volume $\rV(\rA)$ corresponds to the intersection of the ellipses associated with the covariances:  \com{$\rV(\rA_{\mathrm{CI}}) = \rE(\mymatrix{\tilde P}_1) \cap \rE(\mymatrix{\tilde P}_2)$} \cite{julier1997nondivergent, reinhardt2015minimum}. For SCI and ESCI, the volume $\rV(\rA)$ does not have any simple interpretation. However, as $\rA_{\mathrm{ESCI}} \subseteq \rA_{\mathrm{SCI}} \subseteq \rA_{\mathrm{CI}}$, $\rV(\rA_{\mathrm{ESCI}}) \subseteq \rV(\rA_{\mathrm{SCI}}) \subseteq \rV({\rA_\mathrm{CI}})$ as illustrated in Figure~\ref{fig: Minimal volumes}.

\com{Containing the minimal volume $\rV(\rA)$ is only a necessary condition for being a conservative upper bound. To be a conservative upper bound, the matrix $\mymatrix{B}$ should be greater than all the $\mymatrix{K} \mymatrix{P}_\cent \mymatrix{K}^{\intercal}$, which is a stronger requirement. However, if a conservative upper bound \emph{tightly} circumscribes the minimal volume, there cannot be smaller conservative upper bounds. The notion of tightness is introduced in Section~\ref{ssec: Tightness} but first the rest of this section gives an alternative characterization of the volume $\rV(\rA_{\mathrm{ESCI}})$ based on the ESCI bounds}.

As ellipsoids and bounds are defined in terms of precision matrices, it is more convenient to manipulate their inverses. Introduce the precision matrices:
\begin{subequations}
\begin{align}
	\mymatrix{A}_F^{\mathrm{ESCI}}(\omega) &\eqdef \mymatrix{B}^{\mathrm{ESCI}}_F(\omega)^{-1}, \\
	\mymatrix{A}_\cent^{\mathrm{ESCI}}(\omega) &\eqdef \mymatrix{B}^{\mathrm{ESCI}}_\cent(\omega)^{-1}, \\
	\mymatrix{\tilde M}_F^*(\mymatrix{P}_\cent) &\eqdef \mymatrix{\tilde P}^*_F(\mymatrix{P}_\cent)^{-1}.
\end{align}
\end{subequations}
Define also the functions:
\begin{subequations}
\begin{align}
	h_{\myvector{x}} :& \left\{\begin{array}{ccl}
		[0,1] & \longrightarrow & \R\\
		\omega &\longmapsto & \myvector{x}^\intercal \mymatrix{A}_F^\mathrm{ESCI}(\omega) \myvector{x}
	\end{array}\right., \\
	g :& \left\{\begin{array}{ccl}
		\R^d & \longrightarrow & \R\\
		\myvector{x} &\longmapsto &\min\limits_{\mymatrix{P}_{\cent} \in \rA_{\mathrm{ESCI}}} \myvector{x}^\intercal \mymatrix{\tilde M}_F^*(\mymatrix{P}_{\cent}) \myvector{x}
	\end{array}\right..
\end{align}
\end{subequations}
These functions characterize the volumes $\rE(\mymatrix{B}_F^{\mathrm{ESCI}}(\omega))$ and $\rV(\rA_{\mathrm{ESCI}})$. By definition, a point $\myvector{x} \in \R^d$ lies inside the ellipsoids associated with a covariance $\mymatrix{P}$ if and only if $\myvector{x}^{\intercal}\mymatrix{P}^{-1}\myvector{x} \le 1$. Therefore:
\begin{subequations}
	\begin{align}
	\rE(\mymatrix{B}_F^{\mathrm{ESCI}}(\omega)) &= \set{\myvector{x} \suchthat h_{\myvector{x}}(\omega) \le 1}, \\
	\rV(\rA_{\mathrm{ESCI}}) &= \set{\myvector{x} \suchthat g(\myvector{x}) \le 1}. \label{eq: Characterisation volume g}
\end{align}
\end{subequations}
As the ESCI bounds are conservative, Lemma~\ref{lem: Inequality optimal bounds} implies that:
\begin{subequations}
\begin{align}
	\forall \omega \in [0,1],\, & \forall \mymatrix{P}_\cent \in \rA_{\mathrm{ESCI}}, & \mymatrix{A}_{F}^{ESCI} &\preceq \mymatrix{\tilde M}_F^*, \\
	\forall \omega \in [0,1],\, & \forall \myvector{x} \in \R^d, & h_{\myvector{x}}(\omega) &\le g(\myvector{x}). \label{eq: Inegalité h et g}
\end{align}
\end{subequations}
Finally, the two following lemmas are required for the proof of our main result.

\begin{lem}\label{lem: Concavity}
	Under Assumption~\ref{as: Positivity}, for all $\myvector{x} \ne \myvector{0}$, $h_{\myvector{x}}$ is strictly concave on $[0,1]$.
\end{lem}
\begin{proof} \com{Provided in Appendix~\ref{proof: lem: Concavity}.}
\end{proof}
\begin{lem}\label{lem: Correlation matrices}
	Let $\mymatrix{\Omega} \in \R^{n\times n}$ be a matrix and define:
	\begin{subequations}\label{eq: Correlation matrix}
		\begin{align}
			\mymatrix{P}_{1,2}^{(1)}(\mymatrix{\Omega}) &= \left(\mymatrix{\tilde P}_1^{(1)}\right)^{1/2}\mymatrix{\Omega}\left(\mymatrix{\tilde P}_2^{(1)}\right)^{1/2} \\
			\mymatrix{P}_\cent(\mymatrix{\Omega}) &\eqdef \begin{bmatrix}
				\mymatrix{\tilde P}_1^{(1)} & \mymatrix{P}_{1,2}^{(1)}(\mymatrix{\Omega}) \\
				\mymatrix{P}_{1,2}^{(1)}(\mymatrix{\Omega})^{\intercal} & \mymatrix{\tilde P}_2^{(1)}
			\end{bmatrix} + \mymatrix{\tilde P}_\cent^{(2)}
		\end{align}
	\end{subequations}
	If $\mymatrix{\Omega}^\intercal\mymatrix{\Omega} \preceq \mymatrix{I}$, then $\mymatrix{P}_\cent(\mymatrix{{\Omega}}) \in \rA_{\mathrm{ESCI}}$.
\end{lem}
\begin{proof} \com{Provided in Appendix~\ref{proof: lem: Correlation matrices}.}
\end{proof}

With these notations and these two lemmas, we are in a position to characterize function $g$ or equivalently the volume $\rV(\rA_{\mathrm{ESCI}})$.
\begin{thm}\label{the: Expression of g}
	Let $\myvector{x}\in \R^d$, $\myvector{x} \ne \myvector{0}$. Under Assumption~\ref{as: Positivity}, the three following cases are mutually exclusive and collectively exhaustive.
	\begin{enumerate}
		\item $h_{\myvector{x}}'(0) < 0$. In this case, $g(\myvector{x}) = h_{\myvector{x}}(0)$.
		\item $h_{\myvector{x}}'(1) > 0$. In this case, $g(\myvector{x}) = h_{\myvector{x}}(1)$.
		\item There exists a unique $\omega_0 \in [0,1]$ such that $h_{\myvector{x}}'(\omega_0) = 0$. In this case, $g(\myvector{x}) = h_{\myvector{x}}(\omega_0)$.
	\end{enumerate}
\end{thm}
\begin{proof}
	Let $\myvector{x}\in \R^d$, $\myvector{x} \ne \myvector{0}$ be set.
	As $\myvector{x} \neq \mymatrix{0}$, according to Lemma~\ref{lem: Concavity}, $h_{\myvector{x}}$ is strictly concave on $[0,1]$. Therefore, the three cases are mutually exclusive and collectively exhaustive.

	To harmonize the notations, let $\omega_0 = 0$ in the first case and $\omega_0 = 1$ in the second case, we need to prove that for the all cases $g(\myvector{x}) = h_{\myvector{x}}(\omega_0)$. By \eqref{eq: Inegalité h et g}, we already have $h_{\myvector{x}}(\omega_0) \le g(\myvector{x})$. Assume for a moment that there exists some $\mymatrix{P}_{\cent}^* \in \rA_{\mathrm{ESCI}}$ such that $\myvector{x}^\intercal \mymatrix{\tilde M}_F^*(\mymatrix{P}_{\cent}^*)\myvector{x} = h_{\myvector{x}}(\omega_0)$. Then, we would also have $g(\myvector{x}) \le h_{\myvector{x}}(\omega_0)$. Therefore, let us prove for each case that there exists a $\mymatrix{P}_{\cent}^* \in \rA_{\mathrm{ESCI}}$ such that $\myvector{x}^\intercal \mymatrix{\tilde M}_F^*(\mymatrix{P}_{\cent}^*)\myvector{x} = h_{\myvector{x}}(\omega_0)$.
	
	Before starting the exhaustion of cases, let us decompose the bound $\mymatrix{B}_\cent^{\mathrm{ESCI}}(\omega) = \mymatrix{\bar{B}}_\cent \mymatrix{D}(\omega)^{-1}$ with:
	\begin{subequations}
		\begin{align*}
			\mymatrix{\bar{B}}_\cent &\eqdef \begin{bmatrix}
				\mymatrix{\tilde P}_1^{(1)} + \omega \mymatrix{\tilde P}_1^{(2)} & \bar\omega \mymatrix{\tilde P}_{1,2}^{(2)} \\
				\omega \mymatrix{\tilde P}_{2,1}^{(2)} & \mymatrix{P}_2^{(1)} + \bar\omega \mymatrix{\tilde P}_2^{(2)}
			\end{bmatrix}, \\
			\mymatrix{D}(\omega) &\eqdef \diag(\omega \mymatrix{I}_d, \bar\omega\mymatrix{I}_d).
		\end{align*}
	\end{subequations}
	With these notations and by letting $\mymatrix{\bar{A}}_\cent \eqdef \mymatrix{\bar{B}}_\cent^{-1}$, the derivative of $h_{\myvector{x}}$ is:
	\begin{align*}
		h'_{\myvector{x}}(\omega) &= - \myvector{x}^{\intercal}\mymatrix{H}^{\intercal}\mymatrix{A}_\cent^{\mathrm{ESCI}}(\omega)\mymatrix{B}_\cent^{\mathrm{ESCI}\prime}(\omega)\mymatrix{A}_\cent^{\mathrm{ESCI}}(\omega)\mymatrix{H}\myvector{x}, \\
		&= \myvector{x}^{\intercal}\mymatrix{H}^{\intercal}\mymatrix{\bar{A}}_{\cent}(\omega)^{\intercal}\begin{bmatrix}
			\mymatrix{\tilde P}_1^{(1)} & \mymatrix{0} \\
			\mymatrix{0} & -\mymatrix{\tilde P}_2^{(1)}
		\end{bmatrix} \mymatrix{\bar{A}}_{\cent}(\omega) \mymatrix{H}\mymatrix{x}.
	\end{align*}
	Introducing the vector $\myvector{y} = \begin{pmatrix} \myvector{y}_1^{\intercal} & \myvector{y}_2^{\intercal}	\end{pmatrix}^{\intercal} = \mymatrix{\bar{A}}_{\cent}(\omega_0) \mymatrix{H}\mymatrix{x} \in \R^{2d}$, we have:
	\begin{equation}\label{eq: Proof derivative h}
		h_{\myvector{x}}'(\omega_0) = \myvector{y}_1^{\intercal} \mymatrix{\tilde P}_1^{(1)} \myvector{y}_1 - \myvector{y}_2^{\intercal} \mymatrix{\tilde P}_2^{(1)} \myvector{y}_2.
	\end{equation}
	We are now in a position to start the exhaustion of the cases.
		
	\case{Case 1} Assume $h_{\myvector{x}}'(0) < 0$, by \eqref{eq: Proof derivative h}:
	\begin{equation*}
		\myvector{y}_1^{\intercal} \mymatrix{\tilde P}_1^{(1)} \myvector{y}_1 \le \myvector{y}_2^{\intercal} \mymatrix{\tilde P}_2^{(1)} \myvector{y}_2.
	\end{equation*}
	Consider $\mymatrix{P}_\cent^* = \mymatrix{P}_\cent(\mymatrix{{\Omega}})$, given by \eqref{eq: Correlation matrix}, with the correlation matrix:
	\begin{equation*}
		\mymatrix{\Omega} = \frac{\left(\mymatrix{\tilde P}_1^{(1)}\right)^{1/2} \myvector{y}_1 \myvector{y}_2^{\intercal}\left(\mymatrix{\tilde P}_2^{(1)}\right)^{1/2}}{\myvector{y}_2^{\intercal} \mymatrix{\tilde P}_2^{(1)} \myvector{y}_2}
	\end{equation*}
	By construction, the only non-null eigenvalue of $\mymatrix{\Omega}^{\intercal}\mymatrix{\Omega}$ is $(\myvector{y}_1^{\intercal} \mymatrix{\tilde P}_1^{(1)} \myvector{y}_1) / (\myvector{y}_2^{\intercal} \mymatrix{\tilde P}_2^{(1)} \myvector{y}_2) \le 1$; it is associated with the eigenvector
	$(\mymatrix{\tilde P}_2^{(1)})^{1/2}\myvector{y}_2$.
	Therefore, $\mymatrix{\Omega}^{\intercal}\mymatrix{\Omega} \preceq \mymatrix{I}_d$ and, by Lemma~\ref{lem: Correlation matrices}, $\mymatrix{P}_\cent^* \in \rA_{\mathrm{ESCI}}$.
	Let us finally prove that $h_{\myvector{x}}(0) = \myvector{x}^{\intercal}\mymatrix{\tilde M}_F^*(\mymatrix{P}_\cent^*)\myvector{x}$. Using the Woodbury inversion identity, the precision matrix $\mymatrix{\tilde M}_F^*(\mymatrix{P}_\cent^*)$ is:
	\begin{equation*}
		\mymatrix{\tilde M}_F^*(\mymatrix{P}_\cent^*) = \mymatrix{\tilde P}_2^{-1} - (\mymatrix{I}_d - \mymatrix{\tilde P}_2^{-1} \mymatrix{P}_{2,1}^*) \mymatrix{R}_1^{-1} (\mymatrix{I}_d - \mymatrix{P}_{1,2}^*\mymatrix{\tilde P}_2^{-1}),
	\end{equation*}
	with $\mymatrix{R}_1 \eqdef \mymatrix{\tilde P}_1 - \mymatrix{P}_{1,2}^*\mymatrix{\tilde P}_2^{-1} \mymatrix{P}_{2,1}^*$.
	Since $h_{\myvector{x}}(0) = \myvector{x}^{\intercal}\mymatrix{\tilde P}_2^{-1}\myvector{x}$, it is sufficient to prove that $(\mymatrix{I}_d - \mymatrix{P}_{1,2}^*\mymatrix{\tilde P}_2^{-1}) \myvector{x} = \myvector{0}$.
	By construction:
	\begin{align*}
		\mymatrix{P}_{1,2}^* &= \frac{\mymatrix{\tilde P}_1^{(1)} \myvector{y}_1 \myvector{y}_2^{\intercal} \mymatrix{\tilde P}_2^{(1)}}{\myvector{y}_2^{\intercal} \mymatrix{\tilde P}_2^{(1)} \myvector{y}_2} + \mymatrix{\tilde P}^{(2)}_{1,2}, \\
		\myvector{y}_1 &= (\mymatrix{\tilde P}_1^{(1)})^{-1} (\mymatrix{I}_d - \mymatrix{\tilde P}_{1,2}^{(2)} \mymatrix{\tilde P}_2^{-1}) \myvector{x},\\
		\myvector{y}_2 &= \mymatrix{\tilde P}_2^{-1} \myvector{x}.
	\end{align*}
	By reinjecting, we verify the equality $(\mymatrix{I}_d - \mymatrix{P}_{1,2}^* \mymatrix{\tilde P}_2^{-1})\myvector{x} = \myvector{0}$. Therefore, $h_{\myvector{x}}(0) = \myvector{x}^{\intercal}\mymatrix{\tilde M}_F^*(\mymatrix{P}_\cent^*)\myvector{x}$ which concludes the first case.	
	 
	\case{Case 2} The second case is symmetrical.
	
	\case{Case 3} Assume $h'_{\myvector{x}}(\omega_0) = 0$. Denote:
	\begin{equation*}
		\gamma \eqdef \myvector{y}_1 \mymatrix{\tilde P}_1^{(1)} \myvector{y}_1 = \myvector{y}_2 \mymatrix{\tilde P}_2^{(1)} \myvector{y}_2 
	\end{equation*}
	Then, consider the matrix $\mymatrix{P}_\cent^* \eqdef \mymatrix{P}_\cent(\Omega)$ with:
	\begin{equation*}
		\mymatrix{\Omega} = \frac{1}{\gamma} \left(\mymatrix{\tilde P}_1^{(1)}\right)^{1/2} \myvector{y}_1 \myvector{y}_2^{\intercal}\left(\mymatrix{\tilde P}_2^{(1)}\right)^{1/2}.
	\end{equation*}
	With same arguments as in Case~$1$, $\mymatrix{\Omega}\mymatrix{\Omega}^{\intercal} \preceq \mymatrix{I}_d$ and $\mymatrix{P}_\cent^* \in \rA_{\mathrm{ESCI}}$.
	Let us finally prove that $h_{\myvector{x}}(\omega_0) = \myvector{x}^{\intercal}\mymatrix{\tilde M}_F^*(\mymatrix{P}_\cent^*)\myvector{x}$. As $\mymatrix{P}_\cent^* \preceq \mymatrix{B}_\cent^{\mathrm{ESCI}}(\omega_0)$, by left and right multiplying by $\mymatrix{A}_\cent^{\mathrm{ESCI}}(\omega_0)$:
	\begin{subequations}
	\begin{equation}\label{eq: Proof first condition inequality}
		\mymatrix{A}_\cent^{\mathrm{ESCI}}(\omega_0)\mymatrix{P}_\cent^* \mymatrix{A}_\cent^{\mathrm{ESCI}}(\omega_0) \preceq \mymatrix{A}_\cent^{\mathrm{ESCI}}(\omega_0).
	\end{equation}
	Furthermore, we claim that:
	\begin{equation}\label{eq: Proof second condition equality}
		(\mymatrix{H}\myvector{x})^{\intercal}\left(\mymatrix{A}_\cent^{\mathrm{ESCI}}(\omega_0)\mymatrix{P}_\cent^* \mymatrix{A}_\cent^{\mathrm{ESCI}}(\omega_0) - \mymatrix{A}_\cent^{\mathrm{ESCI}}(\omega_0)\right)\mymatrix{H}\myvector{x} = 0.
	\end{equation}
	\end{subequations}
	Indeed, on the one hand:
	\begin{align*}
		&\myvector{x}^{\intercal}\mymatrix{H}^{\intercal}\mymatrix{A}_\cent^{\mathrm{ESCI}}(\omega_0)\mymatrix{P}_\cent^* \mymatrix{A}_\cent^{\mathrm{ESCI}}(\omega_0)\mymatrix{H}\myvector{x} \\
		&= \myvector{x}^{\intercal}\mymatrix{H}^{\intercal}\mymatrix{\bar{A}}_\cent(\omega_0)^{\intercal}\mymatrix{D}(\omega_0)\mymatrix{P}_\cent^*\mymatrix{D}(\omega_0)\mymatrix{\bar{A}}(\omega_0)\mymatrix{H}\myvector{x}, \\
		&= \myvector{y}^{\intercal}\mymatrix{D}(\omega_0) \mymatrix{P}_\cent^* \mymatrix{D}(\omega_0) \myvector{y}, \\
		&= \gamma(\omega_0^2 + \bar\omega_0^2 + 2\omega_0\bar\omega_0) + \myvector{y}^{\intercal}\mymatrix{D}(\omega_0)\mymatrix{\tilde P}_\cent^{(2)}\mymatrix{D}(\omega_0)\myvector{y},\\
		&= \gamma + \myvector{y}^{\intercal}\mymatrix{D}(\omega_0)\mymatrix{\tilde P}_\cent^{(2)}\mymatrix{D}(\omega_0)\myvector{y}.
 	\end{align*}
 	And on the other hand:
 	\begin{align*}
 		&\myvector{x}^{\intercal}\mymatrix{H}^{\intercal}\mymatrix{A}_\cent^{\mathrm{ESCI}}(\omega_0)\mymatrix{H}\myvector{x} \\
 		&=\myvector{x}^{\intercal}\mymatrix{H}^{\intercal}\mymatrix{A}_\cent^{\mathrm{ESCI}}(\omega_0)\mymatrix{B}_\cent^{\mathrm{ESCI}}(\omega_0)\mymatrix{A}_\cent^{\mathrm{ESCI}}(\omega_0)\mymatrix{H}\myvector{x} \\
 		&= \myvector{x}^{\intercal}\mymatrix{H}^{\intercal}\mymatrix{\bar{A}}_\cent(\omega_0)^{\intercal}\mymatrix{D}(\omega_0)\mymatrix{\bar{B}}(\omega_0)\mymatrix{\bar{A}}(\omega_0)\mymatrix{H}\myvector{x}, \\
 		&= \myvector{y}^{\intercal}\mymatrix{D}(\omega_0)\mymatrix{\bar{B}}(\omega_0) \myvector{y}, \\
 		&= \myvector{y}^{\intercal}\mymatrix{D}(\omega_0)\mymatrix{B}_\cent^{\mathrm{ESCI}}(\omega_0)\mymatrix{D}(\omega_0) \myvector{y},\\
 		&= \gamma + \myvector{y}^{\intercal}\mymatrix{D}(\omega_0)\mymatrix{\tilde P}_\cent^{(2)}\mymatrix{D}(\omega_0)\myvector{y}.
 	\end{align*}
	Combining \eqref{eq: Proof first condition inequality} and \eqref{eq: Proof second condition equality} gives that $\mymatrix{A}_\cent^{\mathrm{ESCI}}(\omega_0)\mymatrix{P}_\cent^* \mymatrix{A}_\cent^{\mathrm{ESCI}}(\omega_0)\mymatrix{H}\myvector{x} = \mymatrix{A}_\cent^{\mathrm{ESCI}}(\omega_0)\mymatrix{H}\myvector{x}$. To conclude, left multiply by $\myvector{x}^{\intercal}\mymatrix{H}^{\intercal}(\mymatrix{P}_\cent^{*})^{-1}\mymatrix{B}_\cent^{\mathrm{ESCI}}(\omega_0) $ to obtain $\myvector{x}^{\intercal} \mymatrix{A}_F^{\mathrm{ESCI}}(\omega_0) \myvector{x} = \myvector{x}^{\intercal}\mymatrix{\tilde M}_F^*(\mymatrix{P}_\cent^*)\myvector{x}$. This concludes Case~$3$ and the proof.
	
\end{proof}

\begin{figure}
	\centering
	\input{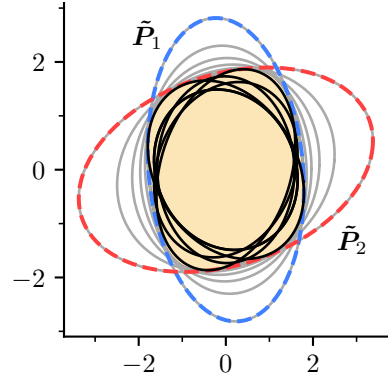}
	\caption{Visualization of equality \eqref{eq: Union and intersection} in the case of SCI. The gray ellipses represent the SCI bounds as in Fig.~\ref{fig: Comparison of the bounds}. The black ellipses represent some particular $\mymatrix{P}_F^*(\mymatrix{P_\cent})$.}
	\label{fig: Union intersection}
\end{figure}

Theorem~\ref{the: Expression of g} has a nice geometric implication. As the ESCI bounds are conservative, for all $\myvector{x}$ and all $\omega \in [0,1]$, $h_{\myvector{x}}(\omega) \le g(\myvector{x})$, so $\max_{\omega}h_{\myvector{x}}(\omega) \le g(\myvector{x})$. According to Theorem~\ref{the: Expression of g}, for all $\myvector{x}$, the upper bound $g(\myvector{x})$ is reached at some $\omega_0$. Consequently, $g$ can be re-expressed as:
\begin{equation*}
	g(\myvector{x}) \eqdef \min\limits_{\mymatrix{P}_{\cent} \in \rA_{\mathrm{ESCI}}} \myvector{x}^\intercal \mymatrix{\tilde M}_F^*(\mymatrix{P}_{\cent}) \myvector{x} =  \max_{\omega \in [0,1]} \myvector{x}^\intercal\mymatrix{A}_F^{\mathrm{ESCI}}(\omega)\myvector{x}.
\end{equation*}
Geometrically, the set $\rV(\rA_{\mathrm{ESCI}})$ can also be re-expressed as:
\begin{equation}\label{eq: Union and intersection}
	\rV(\rA_{\mathrm{ESCI}}) \eqdef \bigcup_{\mymatrix{P}_{\cent} \in \rA_{\mathrm{ESCI}}} \rE(\mymatrix{\tilde P}_F^*(\mymatrix{P}_\cent)) = \bigcap_{\omega \in [0,1]} \rE(\mymatrix{B}_F^{\mathrm{ESCI}}(\omega)).
\end{equation}
Thus, the minimal volume $\rV(\rA_{\mathrm{ESCI}})$ is also characterized by the intersection of the ellipsoids induced by the ESCI bounds. These two equivalent characterizations are  illustrated in Figure~\ref{fig: Union intersection}. As a consequence, $\rV(\rA_{\mathrm{ESCI}})$ is not only a volume common to the all ellipsoids associated with conservative bounds, but it is also the largest volume common to these ellipsoids. 

The common volume $\rV(\rA_{\mathrm{ESCI}})$ of the conservative bounds is now characterized by the ESCI bounds. The next section shows that the ESCI bounds generate also the smallest ellipsoids containing this volume.

\subsection{Tightness of SCI bounds over $\rV(\rA_{\mathrm{ESCI}})$}\label{ssec: Tightness}

First, let us recall the definition of the tightness.
\begin{definition}[Tightness]\label{def: Tightness}
	An ellipsoid $\rE(\mymatrix{P})$ is said to \emph{tightly circumscribe} a volume $\rV$, if:
	\begin{itemize}
		\item[$(i)$] $\rV \subseteq \rE(\mymatrix{P})$ and,
		\item[$(ii)$] for any other ellipsoid $\rE(\mymatrix{Q})$, $\rV \subseteq \rE(\mymatrix{Q}) \subseteq \rE(\mymatrix{P})$ implies $\mymatrix{P} = \mymatrix{Q}$.
	\end{itemize}
\end{definition}
In other words, there is no ellipsoid smaller than $\rE(\mymatrix{P})$ that contains $\rV$. An example of tight ellipsoid is shown in Figure~\ref{fig: Tight bound}.
\begin{figure}
	\centering
	\input{./fig/tight_bound.pgf}
	\caption{Example of tight ellipsoid. The ellipsoid $\rE(\mymatrix{P})$ (solid line) tightly circumscribes $\rV$ (colored area). \com{The ellipsoid $\rE(\mymatrix{Q})$ (dash-dotted line) does not tightly circumscribe $\rV$: for example $\rV \subseteq \rE(\mymatrix{P}) \subseteq \rE(\mymatrix{Q})$.}}
	\label{fig: Tight bound}
\end{figure}

\com{In this section, we prove that the only ellipsoids that tightly circumscribe $\rV(\rA_{\mathrm{ESCI}})$ are ESCI ellipsoids. As a consequence, since all conservative upper bounds must contain $\rV(\rA_{\mathrm{ESCI}})$ there cannot be smaller upper bounds.}

According to Definition~\ref{def: Tightness}, and the characterization of $\rV(\rA_{\mathrm{ESCI}})$ in \eqref{eq: Characterisation volume g}: an ellipsoid $\rE(\myvector{P})$ tightly circumscribes $\rV(\rA_{\mathrm{ESCI}})$ if:
\begin{itemize}
	\item[$(i)$] for all $\myvector{x}$, $\myvector{x}^{\intercal} \mymatrix{P}^{-1} \myvector{x} \le g(\myvector{x})$ and,
	\item[$(ii)$] for all $\mymatrix{Q}$ such that $\myvector{x}^{\intercal} \mymatrix{Q}^{-1} \myvector{x} \le g(\myvector{x})$, $\mymatrix{Q} \preceq \mymatrix{P}$ implies $\mymatrix{Q} = \mymatrix{P}$.
\end{itemize}

The main result is the following theorem. Its proof is inspired by a result of Kahan for the intersection of ellipsoids \cite{kahan1968circumscribing}. It is a proof by exhaustion whose cases have been adapted to ESCI.

\begin{thm}\label{the: Tight inclusion}
	Under Assumption~\ref{as: Positivity}, if $\rE(\mymatrix{B})$ tightly circumscribes $\rV(\rA_{\mathrm{ESCI}})$, then there exists $\omega_1 \in [0,1]$ such that $\mymatrix{B} = \mymatrix{B}_F^{\mathrm{ESCI}}(\omega_1)$.
\end{thm}
\begin{proof}
	Let $\mymatrix{B}$ be a bound whose ellipsoid tightly circumscribes $\rV(\rA_{\mathrm{ESCI}})$ and denote $\mymatrix{A} = \mymatrix{B}^{-1}$ its precision matrix. Let:
	\begin{equation*}
		\phi = \min_{\myvector{x}\ne 0} \frac{g(\myvector{x})}{\myvector{x}^\intercal \mymatrix{A}\myvector{x}}.
	\end{equation*}
	From the definition of $g$:
	\begin{equation*}
		\phi = \min_{\myvector{x}\ne 0} \min_{\mymatrix{P}_{\cent}\in \rA_{\mathrm{ESCI}}} \frac{\myvector{x}^\intercal \mymatrix{\tilde M}_F^*(\mymatrix{P}_{\cent})\myvector{x}}{\myvector{x}^\intercal \mymatrix{A}\myvector{x}}.
	\end{equation*}
	As $\phi$ is the minimum over a compact set (\eg{} $\myvector{x}^\intercal \mymatrix{A} \myvector{x} = 1$), it is achieved at some vector $\myvector{z}$ and some covariance $\mymatrix{P}_{\cent}^*$. Furthermore, by construction $\phi = 1$. Indeed, by conservatism, $\forall \myvector{x}$, $\myvector{x}^\intercal \mymatrix{A}\myvector{x} \le g(\myvector{x})$ so $\phi \ge 1$. And $\phi \le 1$, since otherwise, $\rE(\frac{1}{\phi}\mymatrix{B})$ would be a ellipsoid smaller than $\rE(\mymatrix{B})$ and containing $\rV(\rA_{\mathrm{ESCI}})$, which would contradict the tightness of $\mymatrix{B}$. By Theorem~\ref{the: Expression of g}, there exists a unique $\omega_0 \in [0,1]$ such that $g(\myvector{z}) = \myvector{z}^\intercal \mymatrix{A}_{F}^{\mathrm{ESCI}}(\omega_0)\myvector{z}$. Therefore, $\myvector{z}$ satisfies:
	\begin{equation*}
		\myvector{z}^{\intercal} \mymatrix{A} \myvector{z} = \myvector{z}^{\intercal} \mymatrix{\tilde M}_F^*(\mymatrix{P}_{\cent}^*) \myvector{z} = g(\myvector{z}) = \myvector{z}^\intercal \mymatrix{A}_{F}^{\mathrm{ESCI}}(\omega_0)\myvector{z}.
	\end{equation*}
	
	Assume for the time being, that $\mymatrix{A} \preceq \mymatrix{A}_{F}^{\mathrm{ESCI}}(\omega_0)$, \ie{} $\mymatrix{B}_{F}^{\mathrm{ESCI}}(\omega_0) \preceq \mymatrix{B}$. Using the conservatism of ESCI: $\rV(\rA_{\mathrm{ESCI}}) \subseteq \rE(\mymatrix{B}_F^{\mathrm{ESCI}}(\omega_0))$. So, the tightness of $\mymatrix{B}$ would implies $\mymatrix{B} = \mymatrix{B}_F^{\mathrm{ESCI}}(\omega_0)$, which would conclude the proof. Therefore, to prove the theorem, we will prove that $\mymatrix{A} \preceq \mymatrix{A}_{F}^{\mathrm{ESCI}}(\omega_0)$ by showing that for all $\myvector{y} \in \R^d$:
	\begin{align}\label{eq: (Pr) Inclusion SCI in bound}\tag{$\rC_y$}
		\myvector{y}^\intercal(\mymatrix{A}_{F}^{\mathrm{ESCI}}(\omega_0) - \mymatrix{A})\myvector{y} \ge 0.
	\end{align}
	
	The following result is used several times in the sequel. 
	\begin{lem}\label{lem: (Pr) Main lemma}
		Let $\myvector{y} \in \R^d$, $\eta \in \R$, and define for any $\lambda \in \R$, $\myvector{x}(\lambda) = \eta \myvector{z} + \lambda \myvector{y}$. If one of the two following statements is true, 
		\begin{enumerate}
			\item $g(\myvector{x}(\lambda)) = \myvector{x}(\lambda)^\intercal\mymatrix{A}_{F}^{\mathrm{ESCI}}(\omega_0) \myvector{x}(\lambda)$, for some $\lambda \ne 0$;
			\item $g(\myvector{x}(\lambda)) = \myvector{x}(\lambda)^\intercal\mymatrix{A}_{F}^{\mathrm{ESCI}}(\omega_0) \myvector{x}(\lambda) + o(\lambda^2)$;
		\end{enumerate} 
		then \eqref{eq: (Pr) Inclusion SCI in bound} holds for $\myvector{y}$.
	\end{lem}
	\begin{proof} \com{Provided in Appendix~\ref{proof: lem: (Pr) Main lemma}.}
	\end{proof}
	
	There are three cases to consider to prove Theorem~\ref{the: Tight inclusion}, these are the similar as in Theorem~\ref{the: Expression of g}.
	
	\case{Case 1} Assume $h_{\myvector{z}}'(0) < 0$. In this case, by Theorem~\ref{the: Expression of g}, $\omega_0 = 0$. Let $\myvector{y} \in \R^d$ be set and let us prove \eqref{eq: (Pr) Inclusion SCI in bound}. By continuity of the function $\myvector{x} \mapsto \myvector{x}^\intercal \mymatrix{A}_{F}^{\mathrm{ESCI}\prime}(0)\myvector{x}$, if $\myvector{z}$ is slightly perturbed in the direction $\myvector{y}$, the inequality $\myvector{x}^\intercal\mymatrix{A}_{F}^{\mathrm{ESCI}\prime}(0)\myvector{x} < 0$ still holds. Formally, there exists $\lambda > 0$ such that the vector $\myvector{x} = \myvector{z} + \lambda \myvector{y}$ also satisfies:
	\begin{equation*}
		\myvector{x}^\intercal \mymatrix{A}_{F}^{\mathrm{ESCI}\prime}(0)\myvector{x} < 0.
	\end{equation*}
	Then, Theorem~\ref{the: Expression of g} gives $g(\myvector{x}) = \myvector{x}^\intercal \mymatrix{A}_{F}^{\mathrm{ESCI}}(0)\myvector{x}$. Hence, by applying Lemma~\ref{lem: (Pr) Main lemma}, \eqref{eq: (Pr) Inclusion SCI in bound} holds. Thus, for all $\myvector{y} \in \R^d$, \eqref{eq: (Pr) Inclusion SCI in bound} holds, which concludes the proof for Case 1.
	
	\case{Case 2} Assume $h_{\myvector{z}}'(1) > 0$. In this case, by Theorem~\ref{the: Expression of g}, $\omega_0 = 1$. This case is symmetrical with Case 1.
	
	\case{Case 3}
	Assume finally that $h_{\myvector{z}}'(0) \ge 0$ and $h_{\myvector{z}}'(1) \le 0$. In this case, by Theorem~\ref{the: Expression of g}, $h_{\myvector{z}}'(\omega_0) = \myvector{z}^\intercal \mymatrix{A}_{F}^{\mathrm{ESCI}\prime}(\omega_0)\myvector{z} = 0$. There are two sub-cases to consider depending on whether $\mymatrix{A}_{F}^{\mathrm{ESCI}\prime}(\omega_0)\myvector{z} = 0$ or not.
	If $\mymatrix{A}_{F}^{\mathrm{ESCI}\prime}(\omega_0)\myvector{z} \ne 0$, we can project \emph{almost} every $\myvector{y}$ to create a vector $\myvector{x}$ that satisfies the first assumption of Lemma~\ref{lem: (Pr) Main lemma}. If $\mymatrix{A}_{F}^{\mathrm{ESCI}\prime}(\omega_0)\myvector{z} = 0$, we cannot, but in this case $g$ and the function $\myvector{x} \mapsto \myvector{x}^\intercal\mymatrix{A}_{F}^{\mathrm{ESCI}}(\omega_0)\myvector{x}$ \emph{coincide} in a neighborhood of $\myvector{z}$ and we can apply the second case of Lemma~\ref{lem: (Pr) Main lemma}.
	
	\case{Case 3.1} Assume that $\mymatrix{A}_{F}^{\mathrm{ESCI}\prime}(\omega_0) \myvector{z} \ne \myvector{0}$. Let $\myvector{y} \in \R^d$ such that $\myvector{y}^\intercal \mymatrix{A}_{F}^{\mathrm{ESCI}\prime}(\omega_0)\myvector{z} \ne 0$. Define:
	\begin{equation*}
		\eta = -\frac{1}{2}\frac{\myvector{y}^\intercal \mymatrix{A}_{F}^{\mathrm{ESCI}\prime}(\omega_0)\myvector{y}}{\myvector{y}^\intercal \mymatrix{A}_{F}^{\mathrm{ESCI}\prime}(\omega_0)\myvector{z}},
	\end{equation*}
	so that the vector $\myvector{x} = \eta\myvector{z} + \myvector{y}$ satisfies $\myvector{x}^\intercal \mymatrix{A}_{F}^{\mathrm{ESCI}\prime}(\omega_0)\myvector{x} = 0$. By Theorem~\ref{the: Expression of g}, $g(\myvector{x}) = \myvector{x}^\intercal \mymatrix{A}_{F}^{\mathrm{ESCI}}(\omega_0)\myvector{x}$, then by Lemma~\ref{lem: (Pr) Main lemma}, \eqref{eq: (Pr) Inclusion SCI in bound} holds for $\myvector{y}$. Thus, \eqref{eq: (Pr) Inclusion SCI in bound} holds for all $\myvector{y}$ except those on the hyperplane $\set{\myvector{y} \suchthat \myvector{y}^\intercal\mymatrix{A}_{F}^{\mathrm{ESCI}\prime}(\omega_0)\myvector{z} = 0}$. By continuity of the function $\myvector{x} \mapsto \myvector{x}^\intercal(\mymatrix{A}_{F}^{\mathrm{ESCI}}(\omega_0) - \mymatrix{A})\myvector{x}$, \eqref{eq: (Pr) Inclusion SCI in bound} holds for all $\myvector{y} \in \R^d$. This concludes the proof of Case 3.1.
	
	\case{Case 3.2} Assume that $\mymatrix{A}_{F}^{\mathrm{ESCI}\prime}(\omega_0) \myvector{z} = \myvector{0}$. Let $\myvector{y} \in \R^d$ be set, and define for any $\lambda \in \R$ the vector $\myvector{x}(\lambda) = \myvector{z} + \lambda \myvector{y}$.
	Then, consider the function:
	\begin{equation*}
		\xi : (\lambda, \chi) \mapsto \myvector{x}(\lambda)^\intercal \mymatrix{A}_{F}^{\mathrm{ESCI}\prime}(\chi)\myvector{x}(\lambda).
	\end{equation*}
	It is regular, satisfies $\xi(0, \omega_0) = 0$, and Lemma~\ref{lem: Concavity} gives that $\frac{\partial \xi}{\partial \chi}(0,\omega_0)= \myvector{z}^\intercal \mymatrix{A}_{F}^{\mathrm{ESCI}\prime\prime}(\omega_0)\myvector{z} < 0$. Then, the Implicit Function Theorem, see \eg{} \cite[Theorem 1.3.1]{krantz2002implicit}, states that there exists a continuous and differentiable function $\chi: \lambda \mapsto \chi(\lambda)$ defined on some neighborhood of $\lambda = 0$ such that for all $\lambda$ in that neighborhood:
	\begin{align*}
		\xi(\lambda, \chi(\lambda)) &= \xi(0, \omega_0) = 0, & \chi'(\lambda) = - \frac{\frac{\partial \xi}{\partial \lambda}(\lambda, \chi(\lambda))}{\frac{\partial \xi}{\partial \chi}(\lambda, \chi(\lambda))}.
	\end{align*}
	If $\chi(\lambda) \in [0,1]$, Theorem~\ref{the: Expression of g} implies that $g(\myvector{x}(\lambda)) = \myvector{x}(\lambda)^\intercal\mymatrix{A}_{F}^{\mathrm{ESCI}}(\chi(\lambda))\myvector{x}(\lambda)$.
	Let us therefore consider the function:
	\begin{equation*}
		g_y : \lambda \mapsto \myvector{x}(\lambda)^\intercal \mymatrix{A}_{F}^{\mathrm{ESCI}}(\chi(\lambda))\myvector{x}(\lambda).
	\end{equation*}	
	This function is twice differentiable at $0$, and using the fact that the derivative of $\chi$ at $\lambda = 0$ is:
	\begin{equation*}
		\chi'(0) = - \frac{\frac{\partial \xi}{\partial \lambda}(0, \omega_0)}{\frac{\partial \xi}{\partial \chi}(0, \omega_0)} =-\frac{2\myvector{y}^\intercal\mymatrix{A}_{F}^{\mathrm{ESCI}\prime}(\omega)\myvector{z}}{\myvector{z}^\intercal\mymatrix{A}_{F}^{\mathrm{ESCI}\prime\prime}(\omega)\myvector{z}} = 0,
	\end{equation*}
	we verify that:
	\begin{align*}
		g_y(0) &= \myvector{z}^\intercal \mymatrix{A}_{F}^{\mathrm{ESCI}}(\omega_0)\myvector{z}, & g_y'(0) &= 2\myvector{z}^\intercal \mymatrix{A}_{F}^{\mathrm{ESCI}}(\omega_0)\myvector{y},\\
		g_y''(0) &= 2\myvector{y}^\intercal \mymatrix{A}_{F}^{\mathrm{ESCI}}(\omega_0)\myvector{y}.
	\end{align*}
	Therefore, the series expansion of $g_y$ at $\lambda = 0$ is:
	\begin{align*}
		g_y(\lambda) &= (\myvector{z} + \lambda \myvector{y})^\intercal \mymatrix{A}_{F}^{\mathrm{ESCI}}(\omega_0)(\myvector{z} + \lambda \myvector{y}) + o(\lambda^2),\\
		&= \myvector{x}(\lambda)^\intercal\mymatrix{A}_{F}^{\mathrm{ESCI}}(\omega_0)\myvector{x}(\lambda) + o(\lambda^2). 
	\end{align*}
	If there exists some $\varepsilon > 0$, such that $\forall \lambda \in [-\varepsilon, \varepsilon]$, $\chi(\lambda) \in [0,1]$. Then, by Theorem~\ref{the: Expression of g} on that neighborhood: 
	\begin{align*}
		g(\myvector{x}(\lambda)) &= \myvector{x}(\lambda)\mymatrix{A}_{F}^{\mathrm{ESCI}}(\chi(\lambda))\myvector{x}(\lambda) \\
		&= \myvector{x}(\lambda)\mymatrix{A}_{F}^{\mathrm{ESCI}}(\omega_0)\myvector{x}(\lambda) + o(\lambda^2).
	\end{align*}
	Thus, by Lemma~\ref{lem: (Pr) Main lemma}, \eqref{eq: (Pr) Inclusion SCI in bound} holds.
	On the other hand, if for all $\varepsilon > 0$, there exists $\lambda \in [-\varepsilon, \varepsilon]$ such that $\chi(\lambda) \notin [0,1]$, then necessarily, $\omega_0 = \chi(0) \in \set{0,1}$. Let us assume for example that $\omega_0 = 0$. In that case, for all $\varepsilon$ small enough, there exists $\lambda \in [-\varepsilon, \varepsilon]$, such that $\chi(\lambda) < 0$. Finally, note that:
	\begin{align*}
		&\myvector{x}(\lambda)^\intercal \mymatrix{A}_{F}^{\mathrm{ESCI}\prime}(0) \myvector{x}(\lambda) \\
		&= \myvector{x}(\lambda)^\intercal [\mymatrix{A}_{F}^{\mathrm{ESCI}\prime}(0) - \mymatrix{A}_{F}^{\mathrm{ESCI}\prime}(\chi(\lambda))] \myvector{x}(\lambda),\\
		&= - \chi(\lambda) \myvector{z}^\intercal\mymatrix{A}_{F}^{\mathrm{ESCI}\prime\prime}(0)\myvector{z} + o( \chi(\lambda)).
	\end{align*}
	As Lemma~\ref{lem: Concavity} implies that $\myvector{z}^\intercal\mymatrix{A}_{F}^{\mathrm{ESCI}\prime\prime}(0)\myvector{z} < 0$,
	for $\varepsilon$ small enough, there exists $\lambda \ne 0$ such that $\myvector{x}(\lambda)^\intercal\mymatrix{A}_{F}^{\mathrm{ESCI}\prime}(0)\myvector{x}(\lambda) < 0$. By Theorem~\ref{the: Expression of g}, $g(\myvector{x}(\lambda)) = \myvector{x}(\lambda)^\intercal\mymatrix{A}_{F}^{\mathrm{ESCI}}(0)\myvector{x}(\lambda)$, and Lemma~\ref{lem: (Pr) Main lemma} gives that \eqref{eq: (Pr) Inclusion SCI in bound} holds. The proof of Case 3 and of the theorem is complete.
\end{proof}

\com{The reverse is not always true: a ESCI bound is not necessarily tight over $\rV(\rA_{\mathrm{ESCI}})$.} For example, consider the case where $\mymatrix{\tilde P}_1^{(1)} = \mymatrix{\tilde P}_1^{(2)} = \mymatrix{\tilde P}_2^{(1)} = \mymatrix{\tilde P}_2^{(2)} = \mymatrix{I}_d$ and $\mymatrix{\tilde P}_{1,2}^{(2)} = \mymatrix{0}$. Using the SCI formula, the bound is:
\begin{align*}
	\myvector{B}_F^{\mathrm{SCI}}(\omega) &= \left\{\omega (\mymatrix{I}_d + \omega\mymatrix{I}_d)^{-1} + \bar\omega(\mymatrix{I}_d + \bar\omega \mymatrix{I}_d)^{-1}\right\}^{-1}, \\
	&= \frac{2+\omega\bar\omega}{1+2\omega\bar\omega} \mymatrix{I}_d.
\end{align*}
As the function $\omega \mapsto \frac{2+\omega\bar\omega}{1+2\omega\bar\omega}$ reaches its minimum at $1/2$, $\myvector{B}_F^{\mathrm{SCI}}(1/2)$ is strictly smaller than the other ESCI bounds. Therefore, only $\myvector{B}_F^{\mathrm{SCI}}(1/2)$ can be a tight bound \com{over $\rV(\rA_{\mathrm{ESCI}})$.} However, if $\rE({B}_F^{\mathrm{ESCI}}(\omega))$ \emph{touches} $\rV(\rA_{\mathrm{ESCI}})$, then the bound is tight as claimed in the following theorem.
\begin{thm}\label{the: Condition for tightness}
	Let $\omega_1 \in [0,1]$. Under Assumption~\ref{as: Positivity}, $\rE({B}_F^{\mathrm{ESCI}}(\omega_1))$ tightly circumscribes $\rV(\rA_{\mathrm{ESCI}})$ if and only if there exists $\myvector{x} \ne 0$ such that $g(\myvector{x}) = \myvector{x}^\intercal\mymatrix{A}_{F}^{\mathrm{ESCI}}(\omega_1)\myvector{x}$.
\end{thm}
\begin{proof}
	As the ESCI bounds are conservative, for all $\myvector{x}$, $\myvector{x}^\intercal\mymatrix{A}_{F}^{\mathrm{ESCI}}(\omega_1)\myvector{x} \le g(\myvector{x})$. If $\forall \myvector{x} \ne 0$, $\myvector{x}^\intercal\mymatrix{A}_{F}^{\mathrm{ESCI}}(\omega_1)\myvector{x} < g(\myvector{x})$, consider $\phi = \min_{\myvector{x} \ne 0} g(\myvector{x})/\myvector{x}^\intercal\mymatrix{A}_{F}^{\mathrm{ESCI}}(\omega_1)\myvector{x}$. As $\phi$ is the minimum over a compact set, it is reached at some vector $\myvector{z} \ne 0$. Since $\myvector{z}^\intercal\mymatrix{A}_{F}^{\mathrm{ESCI}}(\omega)\myvector{z} < g(\myvector{z})$, $\phi > 1$. So $\rV(\rA_{\mathrm{ESCI}}) \subseteq \rE(\frac{1}{\phi}{B}_F^{\mathrm{ESCI}}(\omega_1)) \subset  \rE({B}_F^{\mathrm{ESCI}}(\omega_1))$, and $\rE(\mymatrix{B}_F^{\mathrm{ESCI}}(\omega_1))$ does not tightly circumscribe $\rV(\rA_{\mathrm{ESCI}})$.
	
	Conversely, assume that there exists $\myvector{x} \ne 0$ such that $g(\myvector{x}) = \myvector{x}^\intercal\mymatrix{A}_{F}^{\mathrm{ESCI}}(\omega_1)\myvector{x}$ and consider another ellipsoid $\rE(\mymatrix{B})$ such that $\rV(\rA_{\mathrm{ESCI}}) \subseteq \rE(\mymatrix{B}) \subseteq \rE(\mymatrix{B}_F^{\mathrm{ESCI}}(\omega_1))$. We can chose $\rE(\mymatrix{B})$ tight and Theorem~\ref{the: Tight inclusion} states that there exists $\omega_2 \in [0,1]$ such that $\mymatrix{B} = \mymatrix{B}_F^{\mathrm{ESCI}}(\omega_2)$. Let us prove that $\omega_1 = \omega_2$. By assumption $g(\myvector{x})=\myvector{x}^\intercal \mymatrix{A}_{F}^{\mathrm{ESCI}}(\omega_1)\myvector{x}$ and $\myvector{x}^\intercal \mymatrix{A}_{F}^{\mathrm{ESCI}}(\omega_1)\myvector{x} \le \myvector{x}^\intercal \mymatrix{A}_{F}^{\mathrm{ESCI}}(\omega_2)\myvector{x} \le g(\myvector{x})$, so $g(\myvector{x})=\myvector{x}^\intercal \mymatrix{A}_{F}^{\mathrm{ESCI}}(\omega_2)\myvector{x}$. As the function $h_{\myvector{x}} : \omega \mapsto \myvector{x}^\intercal \mymatrix{A}_{F}^{\mathrm{ESCI}}(\omega)\myvector{x}$ is strictly concave on $[0,1]$ by Lemma~\ref{lem: Concavity}, it reaches its maximum exactly once on $[0,1]$. Furthermore, as ESCI is conservative, $\forall \omega \in[0,1]$, $h_{\myvector{x}}(\omega) \le g(\myvector{x}) = h_{\myvector{x}}(\omega_1) = h_{\myvector{x}}(\omega_2)$. Thus, $\omega_1 = \omega_2$, and $\rE(\mymatrix{B}_F^{\mathrm{ESCI}}(\omega_1))$ tightly circumscribes $\rV(\rA_{\mathrm{ESCI}})$.
\end{proof}

\subsection{Proof of the main result}\label{ssec: Proof of the main result}

Thanks to Theorem~\ref{the: Tight inclusion}, we are now in a position to prove the main result, Theorem~\ref{the: Main result}. \com{Let us start by proving Theorem~\ref{the: Tight inclusion} in the special case of Assumption~\ref{as: Positivity}.}

\begin{proof}[Proof of Theorem~\ref{the: Main result} under Assumption~\ref{as: Positivity}]
	
	Let us denote:
	$$\Omega^* = \arg\min_{{\omega} \in [0,1]} J(\mymatrix{B}_{F}^{\mathrm{ESCI}}({\omega}))$$
	and consider $\omega^* \in \Omega^*$ and a conservative bound $\mymatrix{B}_F$ associated with a fusion gain $\mymatrix{K}$.
	
	Assume for the time being that there exists $\omega_1 \in [0, 1]$ such that $\mymatrix{B}_F^{\mathrm{ESCI}}(\omega_1) \preceq \mymatrix{B}_F$. Then, since the cost function $J$ is increasing, for any conservative bound $\mymatrix{B}_F$:
	\begin{equation*}
		J(\mymatrix{B}_{F}^{\mathrm{ESCI}}({\omega^*})) \le J(\mymatrix{B}_{F}^{\mathrm{ESCI}}({\omega_1})) \le J(\mymatrix{B}_{F}),
	\end{equation*}
	with equality if $\mymatrix{B}_F = \mymatrix{B}_{F}^{\mathrm{ESCI}}({\omega_1})$ and $\omega_1 \in \Omega^*$. Thus, $\mymatrix{B}_F$ minimizes the cost if and only if there exists $\omega_1 \in \Omega^*$ such that $\mymatrix{B}_F = \mymatrix{B}_{F}^{\mathrm{ESCI}}({\omega_1})$ as claimed.
	
	Therefore, to prove Theorem~\ref{the: Main result}, we can simply prove that for any conservative bound $\mymatrix{B}_F$, there exists $\omega_1 \in [0, 1]$ such that $\mymatrix{B}_F^{\mathrm{ESCI}}(\omega_1) \preceq \mymatrix{B}_F$. Under Assumption~\ref{as: Positivity}, Theorem~\ref{the: Tight inclusion} gives this result.
\end{proof}

\com{Without Assumption~\ref{as: Positivity}, a short discussion must be added.}

\begin{proof}[Proof of Theorem~\ref{the: Main result} in the general case]
	Consider a conservative bound $\mymatrix{B}_F$ associated with a fusion gain $\mymatrix{K}$. As before, to prove Theorem~\ref{the: Main result}, let us simply prove that there exists $\omega_1 \in [0, 1]$ such that $\mymatrix{B}_F^{\mathrm{ESCI}}(\omega_1) \preceq \mymatrix{B}_F$.
	
	For all $\varepsilon > 0$, define the set:
	\begin{align}
		\rA_{\mathrm{ESCI}}(\varepsilon) &\eqdef \left\{\mymatrix{P}^{(1)}_\cent + (\mymatrix{\tilde P}^{(2)}_\cent + \varepsilon \mymatrix{I}_{2d}) \suchthat \right.\notag\\
		&\qquad \left.\forall i, \ \mymatrix{P}^{(1)}_{i,i} = \mymatrix{\tilde P}^{(1)}_{i} + \varepsilon \mymatrix{I}_d, \mymatrix{P}^{(1)}_\cent \succeq \mymatrix{0} \right\}.
	\end{align}
	This set ensures the non-singularity of the matrices: the matrices $\mymatrix{\tilde P}^{(1)}_1$, $\mymatrix{\tilde P}^{(1)}_2$ and $\mymatrix{\tilde P}^{(2)}_\cent$ have been replaced by the matrices $\mymatrix{\tilde P}^{(1)}_1 + \varepsilon \mymatrix{I}_d$, $\mymatrix{\tilde P}^{(1)}_2 + \varepsilon \mymatrix{I}_d$ and $\mymatrix{\tilde P}^{(2)}_\cent + \varepsilon \mymatrix{I}_{2d}$. In particular, Assumption~\ref{as: Positivity} holds with $\rA_{\mathrm{ESCI}}(\varepsilon)$ and Theorem~\ref{the: Tight inclusion} applies. Furthermore by construction:
	\begin{equation*}
		\mymatrix{P}_\cent \in \rA_{\mathrm{ESCI}} \Leftrightarrow \mymatrix{P}_\cent^\varepsilon \eqdef \mymatrix{P}_\cent + 2\varepsilon \mymatrix{I}_{2d} \in \rA_{\mathrm{ESCI}}(\varepsilon).
	\end{equation*}
	Therefore, if $\forall \mymatrix{P}_\cent \in \rA_{\mathrm{ESCI}}$, $\mymatrix{K}\mymatrix{P}_\cent \mymatrix{K}^{\intercal} \preceq \mymatrix{B}_F$, then $\forall \mymatrix{P}_\cent^\varepsilon \in \rA_{\mathrm{ESCI}}(\varepsilon)$, $\mymatrix{K}\mymatrix{P}_\cent^\varepsilon \mymatrix{K}^{\intercal} \preceq \mymatrix{B}_F + 2\varepsilon\mymatrix{K}\mymatrix{K}^{\intercal}$. Thus, the matrix $\mymatrix{B}_F + 2\varepsilon\mymatrix{K}\mymatrix{K}^{\intercal}$ is a conservative bound for the fusion induced by $\mymatrix{K}$ over the set $\rA_{\mathrm{ESCI}}(\varepsilon)$ and the ellipsoid $\rE(\mymatrix{B}_F + 2\varepsilon\mymatrix{K}\mymatrix{K}^{\intercal})$ contains the volume $\rV(\rA_{\mathrm{ESCI}}(\varepsilon))$. Define the centralized and fused ESCI bounds associated with the set $\rA_{\mathrm{ESCI}}(\varepsilon)$:
	\begin{subequations}
		\begin{align}
			\mymatrix{B}_\cent^{\mathrm{ESCI}}(\omega, \varepsilon) &\eqdef \mymatrix{B}_\cent^{(1)}(\omega, \varepsilon) + \mymatrix{\tilde P}_\cent^{(2)} + \varepsilon \mymatrix{I}_{2d}, \\
			\mymatrix{B}_\cent^{(1)}(\omega, \varepsilon) &\eqdef \diag\left(\frac{1}{\omega} (\mymatrix{\tilde P}_1^{(1)} + \varepsilon\mymatrix{I}_d), \right.\notag\\
			&\qquad \left. \frac{1}{\bar\omega} (\mymatrix{\tilde P}_2^{(1)} + \varepsilon\mymatrix{I}_d)\right),\\
			\mymatrix{B}^{\mathrm{ESCI}}_F(\omega, \varepsilon) &\eqdef \left(\mymatrix{H}^{\intercal}\mymatrix{B}_\cent^{\mathrm{ESCI}}(\omega, \varepsilon)^{-1} \mymatrix{H}\right)^{-1}.
		\end{align}
	\end{subequations} According to Theorem~\ref{the: Tight inclusion}, there exists $\omega_\varepsilon \in [0,1]$ such that:
	\begin{equation*}
		\mymatrix{B}_F^{\mathrm{ESCI}}(\omega_\varepsilon, \varepsilon) \preceq \mymatrix{B}_F + 2\varepsilon\mymatrix{K}\mymatrix{K}^{\intercal}.
	\end{equation*}
	Since the function $\varepsilon \mapsto \mymatrix{B}^{\mathrm{ESCI}}_F(\omega, \varepsilon)$ is non-decreasing:
	\begin{equation*}
		\mymatrix{B}_F^{\mathrm{ESCI}}(\omega_\varepsilon) = \mymatrix{B}_F^{\mathrm{ESCI}}(\omega_\varepsilon, 0) \preceq \mymatrix{B}_F + 2\varepsilon\mymatrix{K}\mymatrix{K}^{\intercal}.
	\end{equation*}
	Since the function $\omega \mapsto \mymatrix{B}_F^{\mathrm{ESCI}}(\omega)$ is continuous on $[0,1]$, by passing to the limit when $\varepsilon$ tends to $0$, there exists $\omega_1 \in [0,1]$ such that $\mymatrix{B}_F^{\mathrm{ESCI}}(\omega_1) \preceq \mymatrix{B}_F$ as claimed.
\end{proof}

\section{Application to distributed estimation}\label{sec: Application}

To illustrate the advantages of the new ESCI fusion rule over the usual SCI rule, this section presents its application to a distributed estimation problem. The scenario considered is extracted from the original article on SCI \cite{julier2001general}.

\subsection{Model}

\begin{figure}
	\centering
	\fontsize{6pt}{40pt}
	{
		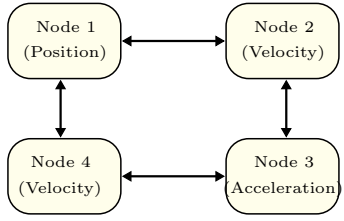%
	}
	\caption{Network considered. The measurement performed by the nodes are given in brackets.}
	\label{fig: Network}
\end{figure}

The sensor network illustrated in Figure~\ref{fig: Network} is used to track the position, velocity, and acceleration of a one-dimensional particle. The network consists of four nodes arranged in a ring structure. Node~$1$ measures the position, Node~$2$ and Node~$4$ measure the velocity, and Node~$3$ measures the acceleration. The dynamics of the particle is assumed to follow the following discrete-time state space model:
\begin{subequations}\label{eq: Dynamic example}
	\begin{align}
		& & \myrandvector{x}(0) &= \myvector{x}_0, \\
		\forall k &\in \N  & \myrandvector{x}(k+1) &= \mymatrix{F}\myrandvector{x}(k) + \myrandvector{w}(k+1).
	\end{align}
\end{subequations}
In \eqref{eq: Dynamic example}, $\myvector{x}_0$ is a given initial state, $\mymatrix{F}$ is the evolution matrix, and $\myrandvector{w}$ is a white noise having covariance $\mymatrix{Q} = \sigma_w^2 \myvector{q}\myvector{q}^{\intercal}$. These matrices are defined as: 
\begin{align*}
	\mymatrix{F} &= \begin{bmatrix}
		1 & \Delta T & \Delta T^2/2 \\
		0 & 1 &\Delta T \\
		0 & 0 & 1
	\end{bmatrix}, &
	\myvector{q} = \begin{bmatrix}
		\Delta T^3/6 \\ \Delta T^2 / 2 \\ \Delta T
	\end{bmatrix}.
\end{align*}
The measurement of Node~$i$ is modeled as:
\begin{equation}\label{eq: Measurment example}
	\myrandvector{z}_i(k) = \mymatrix{H}_i\myvector{x}(k) + \myrand{v}_i(k),
\end{equation}
where $\myvector{H}_i^{\intercal}$ is a vector of the canonical basis, and $\myrand{v}_i(k)$ is an independent white noise whose variance is $\mymatrix{R}_i$. The numerical values used are: $\Delta T = 0.1$~s, $\sigma_w^2 = 100$~(m/s$^3$)$^2$, $\mymatrix{R}_1 = 1$~m$^2$, $\mymatrix{R}_2 = 2$~(m/s)$^2$, $\mymatrix{R}_3 = 0.25$~(m/s$^2$)$^2$ and $\mymatrix{R}_4 = 3$~(m/s)$^2$.

\subsection{Distributed estimation algorithm}

\begin{figure}
	\centering
	\fontsize{6pt}{40pt}
	{
		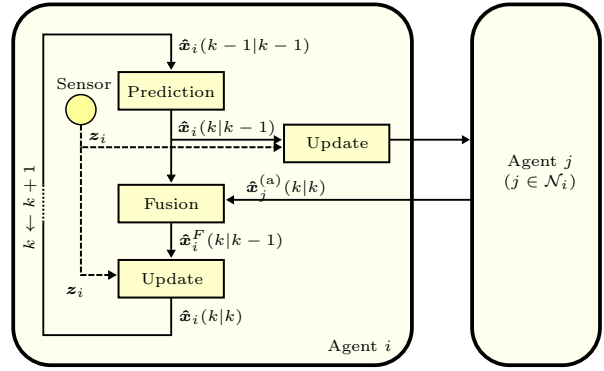%
	}
	\caption{Schematic diagram of the distributed estimation algorithm from the perspective of Node~$i$.}
	\label{fig: Distributed estimation algorithm}
\end{figure}

The four agents apply the same algorithm, also extracted from \cite{julier2001general}. This algorithm is summarized in the diagram in Figure~\ref{fig: Distributed estimation algorithm}. Its four steps are described below from the perspective of Node~$i$.
\begin{enumerate}
	\item Prediction. The state $\myvector{x}(k)$ is predicted from the previous estimator using the evolution model \eqref{eq: Dynamic example}.
	\begin{subequations}
		\begin{align}
		\myrandvector{\hat x}_i(k|k-1) &= \mymatrix{F}\myrandvector{\hat x}(k-1|k-1), \\
		\mymatrix{\tilde P}_i(k|k-1) &= \mymatrix{F}\mymatrix{\tilde P}_i(k-1|k-1)\mymatrix{F^{\intercal} + \myvector{Q}}.
	\end{align}
	\end{subequations}
	\item Update. The estimator is updated using the measurement \eqref{eq: Measurment example}. As the measurement is independent, the correction is made using the information filter.
	\begin{subequations}\label{eq: Update example}
		\begin{align}
		\mymatrix{\tilde P}_i^{(a)}(k|k)^{-1} &= \mymatrix{\tilde P}_i(k|k-1)^{-1} + \myvector{H}_i^{\intercal}\mymatrix{R}_i^{-1} \myvector{H}_i, \\
		\myrandvector{\hat x}^{(a)}_i(k|k) &= \mymatrix{\tilde P}_i^{(a)}(k|k)\mymatrix{\tilde P}_i(k|k-1)^{-1}\myrandvector{\hat x}(k|k-1) \notag \\
		& \qquad + \mymatrix{\tilde P}_i^{(a)}(k|k) \mymatrix{R}_i^{-1}\myvector{H}_i^{\intercal}\myrand{z}_i(k).
	\end{align}
	\end{subequations}
	The supscript $(a)$, for autonomous, indicates that the estimator is built solely using the measurement of Node~$i$. The estimate $\myrandvector{\hat x}^{(a)}_i(k|k)$ is sent to the neighbors $\rN_i$ of Node~$i$, and Node~$i$ also receives the estimates $\myrandvector{\hat x}^{(a)}_j(k|k)$ from its neighbors.
	\item Fusion. The received estimates are fused in a conservative fashion with the prediction $\myrandvector{\hat x}_i(k|k-1)$.
	\begin{equation}
		\set{\begin{array}{c}
			\myrandvector{\hat x}_i(k|k-1),\\
			\myrandvector{\hat x}^{(a)}_j(k|k),\, j \in \rN_i
			\end{array} } \longrightarrow \myrandvector{\hat x}_F(k|k-1).
	\end{equation}
	This fusion may be performed using CI, SCI or ESCI.
	\item Update. The fused estimator is finally updated using the measurement \eqref{eq: Measurment example} as in \eqref{eq: Update example}.
\end{enumerate}

The splitting used to perform the fusion have been introduced in the motivating example of Section~\ref{ssec: Motivating example}.

\subsection{Results and discussion}

\begin{figure*}
	\centering
	\input{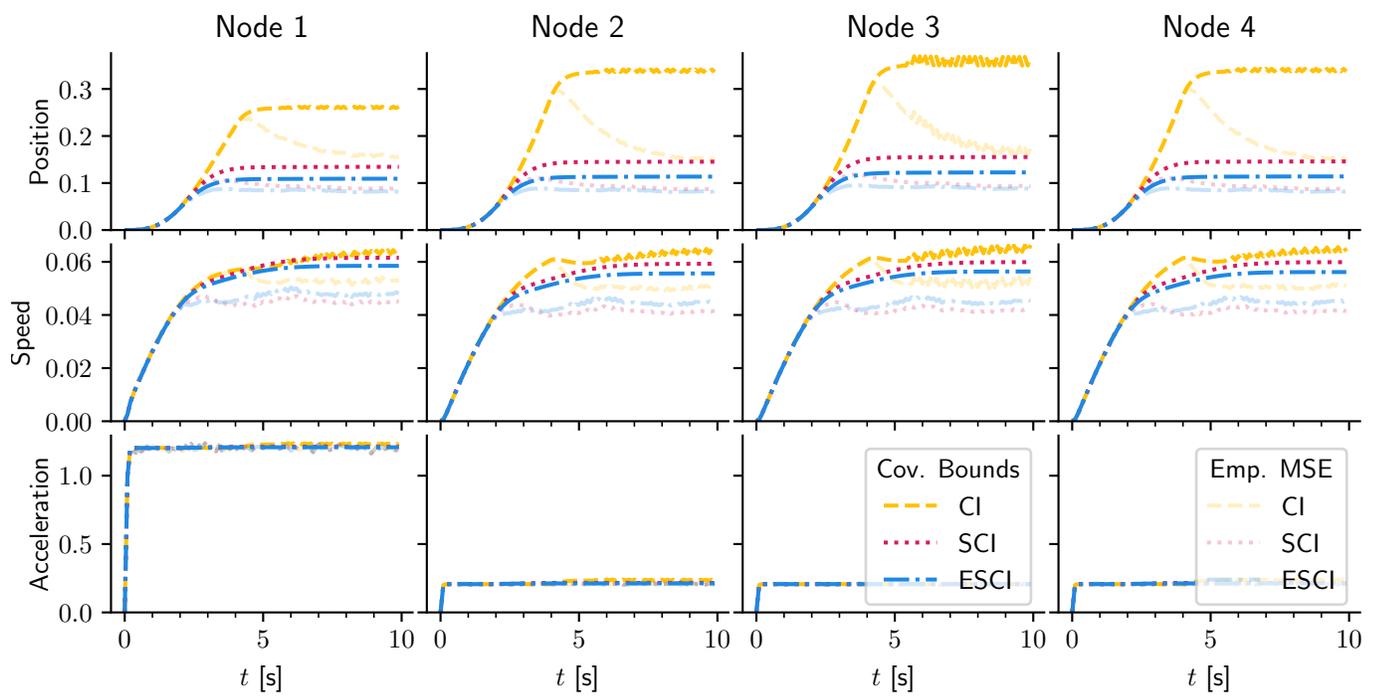}
	\caption{Estimated variance bounds (matte curves) and MSEs (semi-transparent curves) of the estimators of each node.}
	\label{fig: Resultat simulation}
\end{figure*}

$10,000$ trajectory simulations of $10$~s ($100$ iterations) were performed. For each of them, the algorithm was run with the three fusion schemes: CI, SCI and ESCI. The empirical MSE matrices of the estimates were also computed. Figure~\ref{fig: Resultat simulation} shows the evolution of the empirical variances and the conservative upper bounds computed for each fusion scheme. As expected, the empirical variances are lower than the conservative bounds. The ESCI algorithm provides better bounds than the SCI algorithm, which provides better bounds than the CI algorithm. The ESCI bounds are about $20$\% lower than the SCI bounds for position, $5$\% lower for velocity, and $1$\% lower for acceleration.

This toy example showed that using ESCI improves the precision of the estimates. A natural question is what is the cost of this improvement? In the context of distributed estimation, there are two costs to consider: the computational cost and the communication cost. As discussed in Section~\ref{ssec: Particular case of a common noise}, in the context of distributed estimation, since the common components are induced by a common noise, the computational cost of ESCI is $O(Nd^3)$. This is also the cost of SCI, since \eqref{eq: SCI equations} requires the inversion of $N+1$ matrices of size $d$. Thus, these two schemes have equivalent computational costs.

Let us now focus on the communication costs of both fusion schemes. To perform SCI, Node~$i$ needs the estimates of its neighbors $j \in \rN_i$ and the covariances of both components of their errors. Recall that with SCI the error $\myrandvector{\tilde x}_j^{(a)}(k|k)$ is split as:
\begin{align*}
	\myrandvector{\tilde x}_j^{(1)} &= (\mymatrix{I} - \mymatrix{W}_j \mymatrix{H}_j) \myrandvector{\tilde x}_j(k|k-1), &
	\myrandvector{\tilde x}_j^{(2)} &= \mymatrix{W}_j \myrandvector{v}_j(k).
\end{align*}
Therefore, the nodes need to send their estimates $\myrandvector{\tilde x}_j^{(a)}(k|k)$, and two covariance matrices. For example, Node~$j \in \rN_i$ can send the covariance of its errors, $\mymatrix{\tilde P}_j^{(a)}$, and the covariance associated with its measurement: $\mymatrix{\tilde P}^{(m)}_j \eqdef \mymatrix{\tilde P}_j^{(a)}\mymatrix{H}_j^{\intercal}\mymatrix{R}_j^{-1}\mymatrix{H}_j \mymatrix{\tilde P}_j^{(a)}$. When receiving these two matrices, Node~$i$ can deduce the covariances of both terms:
\begin{align*}
	\mymatrix{\tilde P}^{(1)}_j &= \mymatrix{\tilde P}_j^{(a)} - \mymatrix{\tilde P}_j^{(m)}, & \mymatrix{\tilde P}^{(2)}_j &= \mymatrix{\tilde P}_j^{(m)}.
\end{align*}
To perform ESCI, recall that in the presence of a common noise, the errors are split according to \eqref{eq: Error decomposition with common noise}:
\begin{equation*}
	\myrandvector{\tilde x}_j = \myrandvector{\tilde x}_j^{(1)} + \myrandvector{\tilde x}_j^{(\mathrm{ind})} + \mymatrix{M}_j \myrandvector{w}.
\end{equation*}
Node~$i$ needs the estimates of its neighbors $j \in \rN_i$, the covariances of their components $\myrandvector{\tilde x}_j^{(1)}$ and $\myrandvector{\tilde x}_j^{(\mathrm{ind})}$, and the matrix $\mymatrix{M}_j$.
If Node~$j \in \rN_i$ sends the covariances $\mymatrix{\tilde P}_j^{(a)}$ and $\mymatrix{\tilde P}^{(m)}_j$, Node~$i$ can compute all these matrices as:
\begin{align*}
	\mymatrix{M}_j &= \mymatrix{\tilde P}_j^{(a)}(\mymatrix{\tilde P}_j^{(m)})^{-1}, \\
	\mymatrix{\tilde P}_j^{(1)} &= \mymatrix{\tilde P}_j^{(a)} - \mymatrix{\tilde P}_j^{(m)} -  \mymatrix{M}_j\mymatrix{Q}\mymatrix{M}_j^{\intercal}, \\
	\mymatrix{\tilde P}_j^{(\mathrm{ind})} &= \mymatrix{\tilde P}_j^{(m)}.
\end{align*}
Thus, applying the ESCI fusion rule does not require any additional communication cost either.

\section{Discussion}\label{sec: Discussion}

The ESCI fusion rule \eqref{eq: ESCI equations} allows to unify several fusion schemes under a single formalism. ESCI is an extension of SCI, which was already an extension of CI. Consequently, both SCI and CI are special cases of ESCI: SCI corresponds to $\mymatrix{\tilde P}_\cent^{(2)}$ being a block diagonal matrix, and CI corresponds to $\mymatrix{\tilde P}_\cent^{(2)}$ being the zero matrix. Furthermore, ESCI is also a generalization of PCI \cite{ajgl2019rectification}. PCI corresponds to the case where the components $\myrandvector{\tilde x}_i^{(1)}$ and $\myrandvector{\tilde x}_i^{(2)}$ partition the error vector. Finally, ESCI also extends the usual optimal fusion of Lemma~\ref{lem: Optimal linear fusion}. This case corresponds to having no unknown components, \ie{} having all the $\mymatrix{\tilde P}_i^{(1)} = \mymatrix{0}$. The results presented in this paper remain valid for all special cases of ESCI.

From a theoretical point of view, Theorem~\ref{the: Main result} provides a nice justification for the use of ESCI and its derivatives. Let us introduce the optimal ESCI fusion problem.
\begin{prob}[Optimal ESCI Fusion]\label{pro: Simplified problem}
	\begin{equation*}	
		\arg\min\limits_{\myvector{\omega }\in \rK^N}  J(\mymatrix{B}_F^{\mathrm{ESCI}}(\myvector{\omega}))
	\end{equation*}
\end{prob}
The most important implication of Theorem~\ref{the: Main result} is a drastic simplification of Problem~\ref{pro: Main problem} in the case of two estimators. When $N = 2$, solving Problem~\ref{pro: Simplified problem} provides a solution to Problem~\ref{pro: Main problem}. Instead of optimizing for $O(d^2)$ unknowns in Problem~\ref{pro: Main problem}, Problem~\ref{pro: Simplified problem} now only requires optimizing for a single variable lying on a segment. Such an optimization is trivial for modern solvers. Even a linear search would be efficient.
For low dimensions, $d \le 4$, \cite{reinhardt2012closed} proposes closed-form solutions for the parameters $\omega$ that minimize the trace and the determinant of the CI bounds. These solutions may be adapted with ESCI in future work to further speed up the optimization.

Geometrically, the optimality of the ESCI bounds has been proved using the tightness over the minimal set $\rV(\rA_{\mathrm{ESCI}})$. Therefore, the optimality of the SCI fusion does not depend on the (increasing) cost function. However, similar to the CI fusion, the optimal bound generally depends on the cost function: \eg{} optimizing the trace or the determinant generally leads to different optimal bounds. A notable difference with CI is that, except for trivial cases where $\mymatrix{\tilde P}_1 \preceq \mymatrix{\tilde P}_2$ or $\mymatrix{\tilde P}_2 \preceq \mymatrix{\tilde P}_1$, all CI bounds tightly circumscribe the minimal volume \cite{kahan1968circumscribing} (in the case of CI, it is the intersection of the ellipsoids $\rE(\mymatrix{\tilde P}_1)$ and $\rE(\mymatrix{\tilde P}_2)$). This implies that all CI bounds reach the minimal bound for some cost function. This is not the case for ESCI as stated in Theorem~\ref{the: Condition for tightness}.


Unfortunately,  the CI bounds are known to be suboptimal for fusing more than two estimators, see \eg{} \cite{reinhardt2015minimum, ajgl2018fusion}. Consequently, the ESCI bounds are also suboptimal when $N \ge 3$. This means that when combining estimators from more than one ally, the ESCI bounds become suboptimal. Furthermore, even if Node~$i$ has exactly one neighbor $j_1 \in \rN_i$, the fusion bound is optimal only for the first iteration of the algorithm. In fact, after two iterations of the distributed algorithm proposed in Section~\ref{sec: Application}, the fused estimator is implicitly obtained as the fusion of three estimators that are correlated to unknown degrees: $\myrandvector{\hat x}_i(2|1)$,  $\myrandvector{\hat x}_{j_1}^{(a)}(2|1)$ and $\mymatrix{F}\myrandvector{\hat x}_{j_1}^{(a)}(1|0)$. Future work should focus on improving the fusions for more than two estimators.
Although not optimal for $N \ge 3$, ESCI still provides an interesting simplification of the fusion. Optimizing the ESCI bound is an optimization on $O(N)$ variables, compared to $O(N^2 d^2)$ for the general optimization problem, Problem~\ref{pro: Main problem}. For CI and SCI, there are empirical direct techniques to choose $\myvector{\omega}$ without performing this optimization, but at the cost of suboptimality. The adaptation of these techniques for ESCI could also be studied in future work.

When applied to a distributed estimation problem,  ESCI provides tighter conservative bounds than SCI and CI. The improvement of ESCI comes from the fact that it better characterizes the set of admissible covariances $\rA$. ESCI exploits the process noise to reduce this set. Furthermore, in applications where the measurement noises are correlated, ESCI could still be applied (but not SCI). For distributed estimation, ESCI does not require more data transmission or computation than SCI. Finally, the ESCI fusion rule can be applied in other contexts as well. For example, it can be easily integrated into the Diffusion Kalman Filter with CI proposed in \cite{hu2011diffusion}.

\section{Conclusion}\label{sec: Conclusion}

In this paper, the SCI fusion rule has been extended to take advantage of the correlated components in the error. The new ESCI brings a common formalism for different state-of-the-art fusion schemes. The ESCI fusion rule is shown to achieve the minimal covariance bound for the fusion of two estimators. Furthermore, ESCI fits perfectly into the common distributed estimation algorithms, where it exploits the commonly observed process noise. In this context, ESCI provides better guarantees on the estimation error at no additional computational or communication cost compared to SCI.

There are several possible avenues for future work. To speed up the optimization of the bound, existing fast fusion techniques could be adapted to the ESCI context, or exact solutions could be derived for small dimensions. However, the most challenging problem is to derive new conservative fusion schemes that would provide better bounds for the fusion of more than two estimators.

\appendix

\subsection{Proof of Lemma~\ref{lem: Optimal linear fusion}}\label{proof: lem: Optimal linear fusion}

Let $\mymatrix{P}_\cent \succ \mymatrix{0}$, denote $\mymatrix{\tilde P}_F^* \eqdef \mymatrix{\tilde P}_F(\mymatrix{K}^*) = (\mymatrix{H}^{\intercal} \mymatrix{\tilde P}_\cent^{-1}\mymatrix{H})^{-1}$, and consider $\mymatrix{K}$ such that $\mymatrix{K}\mymatrix{H} = \mymatrix{I}_d$.
\begin{align*}
	\mymatrix{0} &\preceq \left[\mymatrix{K} - \mymatrix{K}^*\right] \mymatrix{\tilde P}_\cent \left[\mymatrix{K} - \mymatrix{K}^*\right]^{\intercal}, \\
	&\preceq \left[\mymatrix{K} - \mymatrix{\tilde P}_F^*\mymatrix{H}^{\intercal}\mymatrix{\tilde P}_\cent^{-1}\right] \mymatrix{\tilde P}_\cent \left[\mymatrix{K} - \mymatrix{\tilde P}_F^*\mymatrix{H}^{\intercal}\mymatrix{\tilde P}_\cent^{-1}\right]^{\intercal}, \\
	&\preceq \mymatrix{\tilde P}_F(\mymatrix{K}) - \mymatrix{\tilde P}_F^*.
\end{align*}
Thus, $\mymatrix{\tilde P}_F^* \preceq \mymatrix{\tilde P}_F(\mymatrix{K})$ as claimed.

\subsection{Proof of Lemma~\ref{lem: Centralized bound to fused bound}}\label{proof: lem: Centralized bound to fused bound}

Let $\mymatrix{B}_\cent$ be an upper bound on $\rA$. Define $\mymatrix{B}_F = (\mymatrix{H}^{\intercal}\mymatrix{B}_\cent^{-1}\mymatrix{H})^{-1}$ and $\mymatrix{K} = \mymatrix{B}_F \mymatrix{H}^{\intercal} \mymatrix{B}_\cent^{-1}$.
By construction, for all $\mymatrix{P}_\cent \in \rA$:
\begin{equation*}\mymatrix{\tilde P}_F(\mymatrix{K}, \mymatrix{P}_\cent) = \mymatrix{K}\mymatrix{P}_\cent \mymatrix{K}^{\intercal} \preceq \mymatrix{K}\mymatrix{B}_\cent \mymatrix{K}^{\intercal} = \mymatrix{B}_F.\end{equation*}
Thus, $\mymatrix{B}_F$ is a conservative bound over $\rA$ for the fusion induced by $\mymatrix{K}$.

\subsection{Proof of Lemma~\ref{lem: CI centralized bound}}\label{proof: lem: CI centralized bound}

The proof is performed by induction on the number of estimators involved in the fusion. Let $\mymatrix{P}_\cent \in \rA_{\mathrm{CI}}$, the property to prove for all $n \in \intEnt{1}{N}$ is $\rP(n)$: $\forall \myvector{\omega} \in \interior\rK^n$:
\begin{equation*}
	\begin{bmatrix}
		\mymatrix{P}_{1,1} & \mymatrix{P}_{1,2} & \cdots & \mymatrix{P}_{1,n} \\
		\mymatrix{P}_{2,1} & \mymatrix{P}_{2,2} & \cdots & \mymatrix{P}_{2,n} \\
		\vdots & \vdots & \ddots & \vdots \\
		\mymatrix{P}_{n,1} & \mymatrix{P}_{n,2} & \cdots & \mymatrix{P}_{n,n}
	\end{bmatrix} \preceq \diag\left(\omega_1^{-1} \mymatrix{\tilde P}_1,\, \dots, \omega_n^{-1} \mymatrix{\tilde P}_n\right).
\end{equation*}
$\rP(1)$ is clearly true. Assume $\rP(n-1)$ for $n \ge 2$ and let us prove $\rP(n)$. Let $\myvector{\omega} \in \interior\rK^{n}$, by isolating the first block-row and first block-column in $\mymatrix{P}_\cent$, we claim that:
\begin{equation*}
	\mymatrix{P}_\cent = \begin{bmatrix}
		\mymatrix{P}_{1,1} & \mymatrix{P}_{1, 2:n} \\
		\mymatrix{P}_{2:n, 1} & \mymatrix{P}_{2:n, 2:n}
	\end{bmatrix} \preceq
	\begin{bmatrix}
		\frac{1}{\omega_{1}} \mymatrix{\tilde P}_{1} & \mymatrix{0} \\
		\mymatrix{0} & \frac{1}{1 - \omega_{1}} \mymatrix{P}_{2:n, 2:n}
	\end{bmatrix}
\end{equation*}
Indeed, the difference is positive semi-definite by the Schur criterion \cite[Theorem~7.7.7]{horn2012matrix}:
\begin{align*}
	\Delta_1 &= \frac{1-\omega_{1}}{\omega_{1}} \mymatrix{\tilde P}_{1} \succ \mymatrix{0},\\
	\Delta_2 &= \frac{\omega_{1}}{1-\omega_{1}} \mymatrix{P}_{2:n,2:n} - \mymatrix{P}_{2:n,1} \Delta_1^{-1} \mymatrix{P}_{1,2:n}, \\
	&= \frac{\omega_{1}}{1-\omega_{1}} \left(  \mymatrix{P}_{2:n,2:n} - \mymatrix{P}_{2:n,1} \mymatrix{\tilde P}_1^{-1} \mymatrix{P}_{1,2:n}\right) \succeq \mymatrix{0}.
\end{align*}
Then by using $\rP(n-1)$ on $\mymatrix{P}_{2:n,2:n}$ with the vector $\myvector{\gamma} \in \interior\rK^{n-1}$ whose entries are $\gamma_{i} = \omega_{i+1}/(1-\omega_1)$, we get $\rP(n)$. Thus, the property is true for all $n \ge 1$.

\subsection{Proof of Lemma~\ref{lem: Inequality optimal bounds}}\label{proof: lem: Inequality optimal bounds}

Let $\mymatrix{B}_F$ be a conservative bound over the set $\rA$ for the fusion induced by some gain $\mymatrix{K}$. Let $\mymatrix{P}_\cent \in \rA$. By conservatism: $\mymatrix{\tilde P}_F(\mymatrix{K}, \mymatrix{P}_\cent) \preceq \mymatrix{B}_F$. Furthermore according to Lemma~\ref{lem: Optimal linear fusion}: $\mymatrix{\tilde P}_F^*(\mymatrix{P}_\cent) \preceq \mymatrix{\tilde P}_F(\mymatrix{K}, \mymatrix{P}_\cent)$. Thus, $\mymatrix{\tilde P}_F^*(\mymatrix{P}_\cent) \preceq \mymatrix{B}_F$ as claimed.

\subsection{Proof of Lemma~\ref{lem: Concavity}}\label{proof: lem: Concavity}

Let us prove that for all $\omega \in (0,1)$, $\mymatrix{A}_F^{\mathrm{ESCI}\prime\prime}(\omega) \prec \mymatrix{0}$ which is sufficient. Let $\omega \in (0,1)$ be set.

\com{By linearity $\mymatrix{A}_F^{\mathrm{ESCI}\prime\prime}(\omega) = \mymatrix{H}^{\intercal}\mymatrix{A}_\cent^{\mathrm{ESCI}\prime\prime}(\omega)\mymatrix{H}$, with:
$$\mymatrix{A}_\cent^{\mathrm{ESCI}}(\omega) \eqdef \mymatrix{B}_\cent^{\mathrm{ESCI}}(\omega)^{-1} = \left(\mymatrix{B}_\cent^{(1)}(\omega) + \mymatrix{\tilde P}_\cent^{(2)}\right)^{-1}.$$
By applying using the Woodbury inversion identity:
\begin{equation*}
	\mymatrix{A}_\cent^{\mathrm{ESCI}}(\omega) =  \mymatrix{\tilde M}_\cent^{(2)} - \mymatrix{\tilde M}_\cent^{(2)} \left(\mymatrix{A}_\cent^{(1)}(\omega) + \mymatrix{\tilde M}_\cent^{(2)}\right)^{-1}\mymatrix{\tilde M}_\cent^{(2)}
\end{equation*}
with $\mymatrix{\tilde M}_\cent^{(2)} \eqdef  \left(\mymatrix{\tilde P}_\cent^{(2)}\right)^{-1}$ and $\mymatrix{A}_\cent^{(1)}(\omega) \eqdef \mymatrix{B}_\cent^{(1)}(\omega)^{-1}$. Note that $\mymatrix{A}_\cent^{(1)}(\omega)$ depends linearly on $\omega$:
$$\mymatrix{A}_\cent^{(1)}(\omega) = \diag\left(\omega (\mymatrix{\tilde P}_1^{(1)})^{-1}, (1-\omega)(\mymatrix{\tilde P}_2^{(1)})^{-1}\right).$$
Therefore, the derivatives of $\mymatrix{A}_\cent^{\mathrm{ESCI}}(\omega)$ are:
\begin{align*}
	\mymatrix{A}_\cent^{\mathrm{ESCI}\prime}(\omega) &= \mymatrix{\tilde M}_\cent^{(2)} \left(\mymatrix{A}_\cent^{(1)}(\omega) + \mymatrix{\tilde M}_\cent^{(2)}\right)^{-1}\mymatrix{A}_\cent^{(1)\prime}\\
	&\qquad \left(\mymatrix{A}_\cent^{(1)}(\omega) + \mymatrix{\tilde M}_\cent^{(2)}\right)^{-1}\mymatrix{\tilde M}_\cent^{(2)},  \\
	\mymatrix{A}_\cent^{\mathrm{ESCI}\prime\prime}(\omega) &=  -2 \mymatrix{\tilde M}_\cent^{(2)} \left(\mymatrix{A}_\cent^{(1)}(\omega) + \mymatrix{\tilde M}_\cent^{(2)}\right)^{-1}\mymatrix{A}_\cent^{(1)\prime}\\
	&\qquad \left(\mymatrix{A}_\cent^{(1)}(\omega) + \mymatrix{\tilde M}_\cent^{(2)}\right)^{-1} \\
	&\qquad  \mymatrix{A}_\cent^{(1)\prime} \left(\mymatrix{A}_\cent^{(1)}(\omega) + \mymatrix{\tilde M}_\cent^{(2)}\right)^{-1} \mymatrix{\tilde M}_\cent^{(2)}.  \\
\end{align*}
Since by Assumption~\ref{as: Positivity} $\left(\mymatrix{A}_\cent^{(1)}(\omega) + \mymatrix{\tilde M}_\cent^{(2)}\right)^{-1} \succ \mymatrix{0}$, the matrix $\mymatrix{A}_F^{\mathrm{ESCI}\prime\prime}(\omega) \prec \mymatrix{0}$ as claimed.}

\subsection{Proof of Lemma~\ref{lem: Correlation matrices}}\label{proof: lem: Correlation matrices}

Consider $\mymatrix{\Omega}$ such that $\mymatrix{\Omega}^\intercal\mymatrix{\Omega} \preceq \mymatrix{I}$ and define:
	\begin{align*}
		\mymatrix{P}_{1,2}^{(1)}(\mymatrix{\Omega}) &= \left(\mymatrix{\tilde P}_1^{(1)}\right)^{1/2}\mymatrix{\Omega}\left(\mymatrix{\tilde P}_2^{(1)}\right)^{1/2}, \\
		\mymatrix{P}_\cent^{(1)}(\mymatrix{\Omega}) &\eqdef \begin{bmatrix}
			\mymatrix{\tilde P}_1^{(1)} & \mymatrix{P}_{1,2}^{(1)}(\mymatrix{\Omega}) \\
			\mymatrix{P}_{1,2}^{(1)}(\mymatrix{\Omega})^{\intercal} & \mymatrix{\tilde P}_2^{(1)}
		\end{bmatrix}, \\
		\mymatrix{P}_\cent(\mymatrix{\Omega}) &\eqdef \mymatrix{P}_\cent^{(1)}(\mymatrix{\Omega}) + \mymatrix{\tilde P}_\cent^{(2)}.
	\end{align*}
By definition, $\mymatrix{P}_\cent(\mymatrix{\Omega}) \in \rA_{\mathrm{ESCI}}$ if and only if $\mymatrix{P}_\cent^{(1)}(\mymatrix{\Omega}) \succeq \mymatrix{0}$. Furthermore:
\begin{equation*}
	\mymatrix{P}_\cent^{(1)}(\mymatrix{\Omega}) =
	\begin{bmatrix} \mymatrix{\tilde P}^{(1)\frac{1}{2}}_1 & \mymatrix{0} \\ \mymatrix{0}^\intercal & \mymatrix{\tilde P}^{(1)\frac{1}{2}}_2
	\end{bmatrix}
	\begin{bmatrix} \mymatrix{I} & \mymatrix{\Omega} \\ \mymatrix{\Omega}^\intercal & \mymatrix{I}
	\end{bmatrix}
	\begin{bmatrix} \mymatrix{\tilde P}^{(1)\frac{1}{2}}_1 & \mymatrix{0} \\ \mymatrix{0}^\intercal & \mymatrix{\tilde P}^{(1)\frac{1}{2}}_2
	\end{bmatrix}
\end{equation*}
As $\mymatrix{\Omega}^\intercal\mymatrix{\Omega} \preceq \mymatrix{I}$, the matrix $\begin{bmatrix}
	\mymatrix{I} & \mymatrix{\Omega} \\
	\mymatrix{\Omega}^\intercal & \mymatrix{I}
\end{bmatrix}$ is positive semi-definite, see \eg{} \cite[Lemma 7.7.6]{horn2012matrix}, and thus $\mymatrix{P}_{\cent} \in \rA_{\mathrm{ESCI}}$ as claimed.

\subsection{Proof of Lemma~\ref{lem: (Pr) Main lemma}}\label{proof: lem: (Pr) Main lemma}

Let us first prove that $(\mymatrix{A}_{F}^{\mathrm{ESCI}}(\omega_0) - \mymatrix{A})\myvector{z} = \myvector{0}$.
By definition of $\myvector{z}$ and $\mymatrix{\tilde M}_F^*(\mymatrix{P}_{\cent}^*)$:
\begin{align*}
	\myvector{z}^\intercal (\mymatrix{A} - \mymatrix{\tilde M}_F^*(\mymatrix{P}_{\cent}^*)) \myvector{z} &= 0, \\
	\myvector{z}^\intercal (\mymatrix{A}_{F}^{\mathrm{ESCI}}(\omega_0) - \mymatrix{\tilde M}_F^*(\mymatrix{P}_{\cent}^*)) \myvector{z} &= 0.
\end{align*}
As both $\mymatrix{B}$ and $\mymatrix{B}_F^{\mathrm{ESCI}}(\omega_0)$ are conservative, $\mymatrix{A} - \mymatrix{\tilde M}_F^*(\mymatrix{P}_{\cent}^*) \preceq \mymatrix{0} $ and $\mymatrix{A}_{F}^{\mathrm{ESCI}}(\omega_0) - \mymatrix{\tilde M}_F^*(\mymatrix{P}_{\cent}^*) \preceq \mymatrix{0}$. Therefore $\mymatrix{A}\myvector{z} = \mymatrix{\tilde M}_F^*(\mymatrix{P}_{\cent}^*)\myvector{z} = \mymatrix{A}_{F}^{\mathrm{ESCI}}(\omega_0)\myvector{z}$. Hence $(\mymatrix{A}_{F}^{\mathrm{ESCI}}(\omega_0) - \mymatrix{A})\myvector{z} = \myvector{0}$.

Assume now that for some $\eta \in \R$ and some $\lambda \ne 0$, the vector $\myvector{x} = \eta \myvector{z} + \lambda \myvector{y}$ satisfies $g(\myvector{x}) = \myvector{x}^\intercal\mymatrix{A}_{F}^{\mathrm{ESCI}}(\omega_0)\myvector{x}$. As $\mymatrix{A}$ is conservative, $g(\myvector{x}) \ge \myvector{x}^\intercal\mymatrix{A}\myvector{x}$. Expanding $\myvector{x}^\intercal (\mymatrix{A}_{F}^{\mathrm{ESCI}}(\omega_0) - \mymatrix{A})\myvector{x}$ gives:
\begin{equation*}
	0 \le \myvector{x}^\intercal (\mymatrix{A}_{F}^{\mathrm{ESCI}}(\omega_0) - \mymatrix{A})\myvector{x} = \lambda^2\myvector{y}^\intercal (\mymatrix{A}_{F}^{\mathrm{ESCI}}(\omega_0) - \mymatrix{A})\myvector{y}.
\end{equation*}
Thus, $\myvector{y}^\intercal (\mymatrix{A}_{F}^{\mathrm{ESCI}}(\omega_0) - \mymatrix{A})\myvector{y} \ge 0$.

Similarly, if $g(\myvector{x}(\lambda)) = \myvector{x}(\lambda)^\intercal\mymatrix{A}_{F}^{\mathrm{ESCI}}(\omega_0)\myvector{x}(\lambda) + o(\lambda^2)$, then:
\begin{equation*}
	\myvector{y}^\intercal(\mymatrix{A}_{F}^{\mathrm{ESCI}}(\omega_0) - \mymatrix{A})\myvector{y} = \frac{g(\myvector{x}(\lambda)) - \myvector{x}(\lambda)^\intercal\mymatrix{A}\myvector{x}(\lambda)}{\lambda^2} + o(1).
\end{equation*}
As $\forall \lambda$, $g(\myvector{x}(\lambda)) - \myvector{x}(\lambda)^\intercal\mymatrix{A}\myvector{x}(\lambda) \ge 0$:
\begin{equation*}
	\myvector{y}^\intercal(\mymatrix{A}_{F}^{\mathrm{ESCI}}(\omega_0) - \mymatrix{A})\myvector{y} \ge o(1).
\end{equation*}
Thus, as the left-hand side is constant, by considering the limit when $\lambda$ goes to $0$:
$\myvector{y}^\intercal (\mymatrix{A}_{F}^{\mathrm{ESCI}}(\omega_0) - \mymatrix{A})\myvector{y} \ge 0$ as claimed.


\section*{References}

\bibliographystyle{plain} 
\bibliography{references}

\end{document}